\documentclass[11pt]{article}
\usepackage{epstopdf}
\usepackage{color}
\usepackage{subfigure}
\usepackage{amsmath}
\usepackage{amssymb}
\usepackage{graphicx,color}
\usepackage{cite}
\usepackage{enumerate}
\usepackage{amsthm}% theorems and proofs
\usepackage{amsfonts,mathrsfs}
\usepackage{geometry}
\usepackage{ifthen}
\usepackage{graphicx}
\usepackage{float}
\usepackage{bm}
\usepackage{booktabs}

\begin{document}
\begin{sloppypar}
\title{\bf Quantifying coherence of quantum channels based on the generalized $\bm{\alpha}$-$\bm{z}$-relative R\'{e}nyi entropy}
\vskip0.1in
\author{\small Jiaorui Fan$^1$, Zhaoqi Wu$^1$\thanks{Corresponding author. E-mail:
wuzhaoqi\_conquer@163.com}, Shao-Ming
Fei$^{2,3}$\\
{\small\it  1. Department of Mathematics, Nanchang University,
Nanchang 330031, China}\\
{\small\it  2. School of Mathematical Sciences, Capital Normal University, Beijing 100048,
China}\\
{\small\it  3. Max-Planck-Institute for Mathematics in the Sciences,
04103 Leipzig, Germany}}
\date{}
\maketitle

\noindent {\bf Abstract} {\small }\\
By using the Choi-Jamio{\l}kowski isomorphism, we propose a well-defined coherence measure of quantum channels based on the generalized $\alpha$-$z$-relative R\'{e}nyi entropy. In addition, we present an alternative coherence measure of quantum channels by quantifying the commutativity between the channels and the completely dephasing channels with the generalized $\alpha$-$z$-relative R\'{e}nyi entropy. Some elegant properties of the measures are illustrated in detail. Explicit formulas of these coherence measures are derived for some detailed typical quantum channels.\\


\vskip0.2in

\noindent {\bf 1. Introduction}\\\hspace*{\fill}\\
As a fundamental feature of quantum physics, coherence plays an
essential role in quantum information processing. Based on the
framework of quantifying the coherence of quantum states\cite{BCP},
quantifications of quantum coherence have been extensively studied
in terms of the $l_{1}$-norm\cite{BCP}, relative entropy\cite{BCP},
skew information\cite{BAS,Y}, fidelity\cite{UA,LZY} and generalized
$\alpha$-$z$-relative R\'{e}nyi entropy\cite{ZJF}, with various
applications in quantum entanglement, quantum algorithm, quantum
meteorology and quantum
biology\cite{WZFL,SSD,ZHC,DN1,DN2,NBC,J,PH,L1,LM,OZ,HV,PGAC,MBCP,YMG,BD,L2,MCE}.
Yu, Zhang, Xu and Tong \cite{YZX} have presented an alternative
framework for quantifying coherence.

Quantum channels characterize the general evolutions of quantum systems\cite{NC}.
In recent years, fruitful results have been obtained on studies of quantum
channels\cite{BGN,DGM,BKZ,ZSC,TEZ,WZFW,XWF1,XWF2,XWF3,LS,H1,KCP,LY,CG,SCG,Z,V}.
Datta, Sazim, Pati and Agrawal \cite{DSP1} investigated the
coherence of quantum channels by using the Choi-Jamio{\l}kowski isomorphism.
Xu\cite{X} proposed a framework to quantify the coherence of quantum channels
by using the Choi-Jamio{\l}kowski isomorphism, and defined the $l_{1}$-norm coherence
measure of quantum channels. Based on this framework, some quantifiers of coherence
for quantum channels have been given successively, such as maximum relative
entropy\cite{JYF}, robustness\cite{JYF}, fidelity\cite{WGY}, skew information
and Hellinger distance\cite{XHN}. Luo, Ye and Li\cite{LYL} introduced the
coherence weight of quantum channels to investigate the quantum resource theory
of dynamical coherence. Kong, Wu, Lv, Wang and Fei\cite{KWL} presented an
alternative framework to quantify the coherence of quantum channels.
%Jin\cite{JYF} has introduced a new coherence quantifier for quantum channels via maximum relative entropy, and studied the robustness of coherence for quantum channels. Furthermore, the coherence weight of quantum channels has been introduced to investigate the quantum resource theory of dynamical coherence in\cite{LYL}. Wang, Gao and Yan\cite{WGY} have also proposed the fidelity coherence measure of quantum channels, and calculated the fidelity coherence measure of the general form of unitary channels.

On the other hand, Meznaric, Clark and Datta\cite{MCD} formulated a measure of nonclassicality of a quantum operation, which is defined by quantifying the commutativity between a quantum operation and a completely dephasing operation based on the relative entropy. Fan, Guo and Yang\cite{FGY} studied the commutativity between a channel and a completely dephasing channel based on the trace distance, and quantified the coherence of quantum channels via commutativity.

The paper is organized as follows. In Section $2$, we present the
definition of a coherence measure for quantum channels based on the
generalized $\alpha$-$z$-relative R\'{e}nyi entropy via
Choi-Jamio{\l}kowski isomorphism, and verify that it is a
well-defined coherence measure. In Section $3$, we study the
commutativity between the channels and the completely dephasing
channels based on the generalized $\alpha$-$z$-relative R\'{e}nyi
entropy, and derive several elegant properties. In Section $4$, we
obtain explicit formulas of coherence measures with respect to some
typical channels for above two newly-defined measures.
Finally, we conclude with a summary in Section $5$.

\vskip0.1in

\noindent {\bf 2. Coherence of quantum channels by using Choi-Jamio{\l}kowski isomorphism based on the generalized $\bm{\alpha}$-$\bm{z}$-relative R\'{e}nyi entropy}\\\hspace*{\fill}\\
For two arbitrary quantum states $\rho$, $\sigma$ and $\alpha$, $z$
$\in\mathbb{R}$, the generalized $\alpha$-$z$-relative R\'{e}nyi
entropy is defined by\cite{ZJF},
\begin{equation}\label{eq1}
D_{\alpha,z}(\rho,\sigma)=\frac{f_{\alpha,z}^{\frac{1}{\alpha}}(\rho,\sigma)-1}{\alpha-1},
\end{equation}
where
\begin{equation}\label{eq2}
f_{\alpha,z}(\rho,\sigma)=\mathrm{Tr}\left(\sigma^\frac{1-\alpha}{2z}\rho^{\frac{\alpha}{z}}\sigma^{\frac{1-\alpha}{2z}}\right)^{z}.
\end{equation}

Let $\{|i\rangle\}_{i=1}^{d}$ be a set of orthonormal basis of a $d$-dimensional Hilbert space $H$. The set $\mathcal{I}$ of quantum states is said to be incoherent if all the
density matrices are diagonal in this basis. The quantum
coherence $\mathit{C}_{\alpha,z}(\rho)$ of a quantum state $\rho$
induced by the generalized $\alpha$-$z$-relative R\'{e}nyi entropy,
\begin{equation}\label{eq3}
\mathit{C}_{\alpha,z}(\rho)=\mathop{\mathrm{min}}\limits_{\sigma\in \mathcal{I}}\mathit{D}_{\alpha,z}(\rho,\sigma),
\end{equation}
is a well-defined coherence measure in each of the following cases\cite{ZJF}:\\
(1) $\alpha\in(0,1)$ and $z\geq{\max{\{\alpha,1-\alpha\}}}$;\\
(2) $\alpha\in(1,2]$ and $z=\{1,\frac{\alpha}{2}\}$;\\
(3) $\alpha>{1}$ and $z=\alpha$.\\
It can be found that $\mathit{C}_{\alpha,z}(\rho)$ reduces to $\mathrm{ln2}\cdot\mathit{C}_{r}(\rho)$ and $2\cdot \mathit{C}_{s}(\rho)$ when $z=1$, $\alpha \rightarrow 1$ and $z=1$, $\alpha = \frac{1}{2}$, respectively, where $\mathit{C}_{r}(\rho)$ denotes the relative entropy of coherence\cite{BCP} and $\mathit{C}_{s}(\rho)$ denotes the skew information of coherence\cite{Y}.

Let $H_{A}$ and $H_{B}$ be two Hilbert spaces with dimensions $\vert
A\vert$ and $\vert B\vert$, orthonormal bases $\{|i\rangle\}_{i}$
and $\{|\beta\rangle\}_{\beta}$, respectively. We assume that
$\{|i\rangle\}_{i}$ and $\{|\beta\rangle\}_{\beta}$ are fixed and
adopt the tensor basis $\{|i\beta\rangle\}_{i\beta}$ as the fixed
basis when considering the multipartite system $H_{AB}=H_{A}\otimes
H_{B}$. Denote by $\mathcal{D}(H_{A})$ and $\mathcal{D}(H_{B})$ the
set of all density operators on $H_{A}$ and $H_{B}$, respectively.
Denote by $\mathcal{C}_{AB}$ the set of all channels from $\mathcal{D}(H_{A})$ to $\mathcal{D}(H_{B})$, $\mathcal{SC}_{ABA^{'}B^{'}}$ the set of all superchannels from $\mathcal{C}_{AB}$ to $\mathcal{C}_{A^{'}B^{'}}$, $\mathcal{IC}_{AB}$ the set of incoherent channels in $\mathcal{C}_{AB}$, and $\mathcal{ISC}_{ABA^{'}B^{'}}$ the set of incoherent superchannels in $\mathcal{SC}_{ABA^{'}B^{'}}$. A quantum channel $\phi\in{\mathcal{C}_{AB}}$ is a completely positive trace-preserving (CPTP) map. A coherence measure $\mathit{C}$ of quantum channels should satisfy the following conditions\cite{X}:\\
(a) Faithfulness: $\mathit{C}(\phi)\geq0$ for any $\phi\in{\mathcal{C}_{AB}}$, and $\mathit{C}(\phi)=0$ if and only if $\phi\in{\mathcal{IC}_{AB}}$;\\
(b) Nonincreasing under $\mathcal{ISC}s$: $\mathit{C}(\phi)\geq\mathit{C[\mathrm{\Theta}(\phi)]}$ for any $\Theta\in{\mathcal{ISC}_{ABA^{'}B^{'}}}$;\\
(c) Nonincreasing under $\mathcal{ISC}s$ on average: $\mathit{C}\left(\phi\right)\geq\sum\limits_{m}p_{m}\mathit{C}(\phi_{m})$ for any $\Theta\in{\mathcal{ISC}_{ABA^{'}B^{'}}}$, with $\{K_{m}\}_{m}$ an incoherent expression of $\Theta$, $p_{m}=\frac{\mathrm{Tr}(K_{m}J_{\phi}K_{m}^{\dagger})}{\vert{A^{'}}\vert}$ and $J_{\phi_{m}}=\vert{A^{'}}\vert\frac{K_{m}J_{\phi}K_{m}^{\dagger}}{\mathrm{Tr}(K_{m}J_{\phi}K_{m}^{\dagger})}$;\\
(d) Convexity:
$\mathit{C}\left(\sum\limits_{m}p_{m}\phi_{m}\right)\leq\sum\limits_{m}p_{m}\mathit{C}(\phi_{m})$
for any $\{\phi_{m}\}_{m}\subset\mathcal{C}_{AB}$ and probability $\{p_{m}\}_{m}$.

Following the idea in\cite{YZX}, the authors in \cite{KWL} proposed
an alternative framework for quantifying the coherence of quantum
channels which substitutes (c) and (d) with the following additivity,
\begin{align}\label{eq4}
\mathit{C}(\phi)=p_{1}\mathit{C}(\phi_{1})+p_{2}\mathit{C}(\phi_{2}),
\end{align}
where $p_{1}+p_{2}=1$, $\phi_{1}\in\mathcal{C}_{AB_{1}}$, $\phi_{2}\in\mathcal{C}_{AB_{2}}$, $\phi\in\mathcal{C}_{AB}$, $|B|=|B_{1}|+|B_{2}|$,
and $\phi(|i\rangle\langle\beta|)=p_{1}\phi_{1}(|i\rangle\langle\beta|)\oplus p_{2}\phi_{2}(|i\rangle\langle\beta|)$.

According to Theorem $3$ in \cite{X}, if $\mathit{C}$ is a coherence
measure for quantum states which satisfies (a)-(d), then the
coherence measure of quantum channels is defined as
\begin{equation}\label{eq5}
\mathit{C}(\phi)=\mathit{C}\left(\frac{J_{\phi}}{\vert A\vert}\right),
\end{equation}
where $J_{\phi}$ is the Choi matrix corresponding to $\phi$. For
convenience, we denote $\frac{J_{\phi}}{\vert A\vert}$ by
$\mathit{M}_{\phi}$.

Suppose that the Kraus representation of a quantum channel $\phi$ is
$\phi(\rho)=\sum_n K_n\rho K_n^{\dagger}$. According to Eq.
(\ref{eq2}) in \cite{WGY}, we have
\begin{align*}
\mathit{M}_{\phi}=(\mathbf{Id} \otimes
\phi)|\varphi\rangle\langle\varphi| =\sum_{n}(\mathbb{I} \otimes
K_{n})|\varphi\rangle\langle\varphi| (\mathbb{I} \otimes
K_{n})^{\dagger}.
\end{align*}
Here
$|\varphi\rangle=\frac{1}{\sqrt{|A|}}\sum\limits_{i=0}^{|A|-1}|ii\rangle$
is a maximally entangled state in Hilbert space $H_{A}\otimes H_{A}$
, $\mathbf{Id}$ is the identity channel, and $\mathbb{I}$ is the
identity operator.
 \\\hspace*{\fill}\\
\noindent {\bf Definition 1} The generalized $\alpha$-$z$-relative R\'{e}nyi entropy of two arbitrary quantum channels $\phi$, $\widetilde{\phi}$ $\in\mathcal{C}_{AB}$ is defined as
\begin{equation}\label{eq6}
D_{\alpha,z}(\phi,\widetilde{\phi})
=\frac{f_{\alpha,z}^{\frac{1}{\alpha}}(\mathit{M}_{\phi},\mathit{M}_{\widetilde{\phi}})-1}{\alpha-1}.
\end{equation}
\noindent {\bf Definition 2} The coherence measure of a channel $\phi$ induced by the generalized $\alpha$-$z$-relative R\'{e}nyi entropy is defined by
\begin{equation}\label{eq7}
\mathit{C}_{\alpha,z}(\phi)=\mathop{\mathrm{min}}_{{\widetilde{\phi}}\in \mathcal{IC}_{AB}}D_{\alpha,z}(\phi,\widetilde{\phi})
=\mathop{\mathrm{min}}_{\mathit{M}_{\widetilde{\phi}}\in\mathcal{I}}\frac{f_{\alpha,z}^{\frac{1}{\alpha}}(\mathit{M}_{\phi},\mathit{M}_{\widetilde{\phi}})-1}{\alpha-1}.
\end{equation}
In particular, when $z=1$, $\alpha\in(0,1)\cup(1,2]$, by using the Corollary 2 in\cite{ZJF}, we have
\begin{equation}\label{eq8}
\mathit{C}_{\alpha,1}(\phi)=\frac{\sum\limits_{i,\beta}\langle i\beta|M_\phi^{\alpha}|i\beta\rangle^{\frac{1}{\alpha}}-1}{\alpha-1}.
\end{equation}
$\mathit{C}_{\alpha,1}(\phi)$ reduces to $\mathrm{ln2}\cdot\mathit{C}_{r}(\phi)$ and $2\cdot\mathit{C}_{s}(\phi)$ when $\alpha\rightarrow1$ and $\alpha=\frac{1}{2}$, where $\mathit{C}_{r}(\phi)$ denotes the relative entropy  of coherence of quantum channels and $\mathit{C}_{s}(\phi)$ denotes the skew information of coherence of quantum channels\cite{XHN}.
  %can be further rewritten as $\mathit{C}_{\frac{1}{2},1}(\phi)=2-2\sum\limits_{i,\beta}\langle i\beta|M_\phi^{\frac{1}{2}}|i\beta\rangle^{2}$ when $\alpha=\frac{1}{2}$, i.e., it's equal to $2\cdot\mathit{C}_{s}(\phi)$ with $z=1$ and $\alpha=\frac{1}{2}$, where $\mathit{C}_{s}(\phi)$ denotes the skew information of coherence of quantum channels\cite{XHN}.

\vskip0.1in

\noindent {\bf Theorem 1} $\mathit{C}_{\alpha,z}(\phi)$ defined in Eq. (\ref{eq7}) is a well-defined coherence measure.\\\hspace*{\fill}\\
\textbf{Proof}~
According to Eqs. (\ref{eq2}), (\ref{eq6}) and (\ref{eq7}), $\mathit{C}_{\alpha,z}(\phi)$ can be further rewritten as\\
\[
\mathit{C}_{\alpha,z}(\phi)=
\begin{cases}
\frac{1-\mathop{\mathrm{max}}\limits_{\mathit{M}_{\widetilde{\phi}}
\in{\mathcal{I}}}f_{\alpha,z}^{\frac{1}{\alpha}}
(\mathit{M}_{\phi},\mathit{M}_{\widetilde{\phi}})}{1-\alpha} \quad \ \ & 0<\alpha<1,\\
\frac{\mathop{\mathrm{min}}\limits_{\mathit{M}_{\widetilde{\phi}}
\in{\mathcal{I}}}f_{\alpha,z}^{\frac{1}{\alpha}}
(\mathit{M}_{\phi},\mathit{M}_{\widetilde{\phi}})-1}{\alpha-1}
\quad \ \ & \alpha>1.
\end{cases}
\]
From the Lemma $1$ in \cite{ZJF}, it is easy to see that $\mathit{C}_{\alpha,z}(\phi)\geq0$, and $\mathit{C}_{\alpha,z}(\phi)=0$ if and only if $\phi=\widetilde{\phi}$. Thus, $\mathit{C}_{\alpha,z}(\phi)$ satisfies the condition (a).

When $\alpha>1$, denote $\Theta^{'}=\frac{|A|}{|A^{'}|}\Theta$ with $\Theta\in{\mathcal{ISC_{ABA^{'}B^{'}}}}$. Thus, $J_{\Theta^{'}}$ is a CPTP map. Direct calculation shows that
\begin{align*}
f_{\alpha,z}(J_{\Theta^{'}(\phi)},J_{\Theta^{'}(\widetilde{\phi})})
=&f_{\alpha,z}\left(\frac{|A|}{|A^{'}|}J_{\Theta(\phi)},
\frac{|A|}{|A^{'}|}J_{\Theta(\widetilde{\phi})}\right)\\
=&\frac{|A|}{|A^{'}|}f_{\alpha,z}(J_{\Theta(\phi)},J_{\Theta(\widetilde{\phi})})\\
=&|A|f_{\alpha,z}\left(\frac{J_{\Theta(\phi)}}{|A^{'}|},
\frac{J_{\Theta(\widetilde{\phi})}}{|A^{'}|}\right).
\end{align*}
Utilizing the Lemma $2$ in\cite{ZJF}, we have
$f_{\alpha,z}(J_{\Theta^{'}(\phi)},J_{\Theta^{'}(\widetilde{\phi})})\leq f_{\alpha,z}(J_{\phi},J_{\widetilde{\phi}})$.
Then $D_{\alpha,z}(\Theta(\phi),\Theta(\widetilde{\phi}))\leq D_{\alpha,z}(\phi,\widetilde{\phi})$. Therefore,
\begin{align*}
\mathit{C}_{\alpha,z}(\Theta(\phi))&=\mathop{\mathrm{min}}
\limits_{\widetilde{\phi}\in{\mathcal{IC}_{AB}}}D_{\alpha,z}
(\Theta(\phi),\widetilde{\phi})\\
&\leq \mathop{\mathrm{min}}\limits_{\widetilde{\phi}\in{\mathcal{IC}_{AB}}}D_{\alpha,z}
(\Theta(\phi),\Theta(\widetilde{\phi}))\\
&\leq \mathop{\mathrm{min}}\limits_{\widetilde{\phi}\in{\mathcal{IC}_{AB}}}
D_{\alpha,z}(\phi,\widetilde{\phi})\\
&=\mathit{C}_{\alpha,z}(\phi).
\end{align*}
It can be seen that $\mathit{C}_{\alpha,z}(\Theta(\phi))\leq \mathit{C}_{\alpha,z}(\phi)$ when $\alpha>1$. The case of $0<\alpha<1$ can be easily proved in the same way. Hence, the condition (b) follows immediately.

Next we prove that $\mathit{C}_{\alpha,z}(\phi)$ satisfies Eq.
(\ref{eq4}). Suppose that $M_{\phi}$ is block-diagonal in the
reference $\{|i\beta\rangle\}_{i\beta}$,
\begin{align*}
M_{\phi}=p_{1}M_{\phi_{1}}\oplus p_{2}M_{\phi_{2}},
\end{align*}
where $p_{1}, p_{2}>0$ with $p_{1}+p_{2}=1$, and $M_{\phi_{1}}$ and
$M_{\phi_{2}}$ are the Choi states (density operators) corresponding
to $\phi_{1}$ and $\phi_{2}$. $M_{\widetilde{\phi}}$, the Choi state
corresponding to $\widetilde{\phi}$, can be written as
\begin{align*}
M_{\widetilde{\phi}}=q_{1}M_{\widetilde{\phi}_{1}}\oplus q_{2}M_{\widetilde{\phi}_{2}},
\end{align*}
where $q_{1}, q_{2}>0$ with $q_{1}+q_{2}=1$, and
$M_{\widetilde{\phi}_{1}}$ and $M_{\widetilde{\phi}_{2}}$ are the
Choi states (density operators) corresponding to
$\widetilde{\phi}_{1}$ and $\widetilde{\phi}_{2}$. Denote by
$\Delta$ either max or min. Let
$t_{m}=\Delta_{M_{\widetilde{\phi}_{m}}}\mathrm{Tr}
\left(M_{\widetilde{\phi}_{m}}^{\frac{1-\alpha}{2z}}
M_{{\phi}_{m}}^{\frac{\alpha}{z}}
M_{\widetilde{\phi}_{m}}^{\frac{1-\alpha}{2z}}\right)^{z}$, $m= 1,
2$. It can be derived that
\begin{align*}
\Delta_{M_{\widetilde{\phi}}\in\mathcal{I}}\mathrm{Tr}\left(
M_{\widetilde{\phi}}^{\frac{1-\alpha}{2z}}
M_{\phi}^{\frac{\alpha}{z}}
M_{\widetilde{\phi}}^{\frac{1-\alpha}{2z}}
\right)^{z}
=\Delta_{q_{1},q_{2}}(q_{1}^{1-\alpha}p_{1}^{\alpha}t_{1}+q_{1}^{1-\alpha}p_{2}^{\alpha}t_{2}).
\end{align*}
Using the H\"{o}lder inequality with $0<\alpha<1$, we have
\begin{align*}
q_{1}^{1-\alpha}p_{1}^{\alpha}t_{1}+q_{2}^{1-\alpha}p_{2}^{\alpha}t_{2}\leq\left(\sum_{m=1,2}p_{m}t_{m}^{\frac{1}{\alpha}}\right)^{\alpha},
\end{align*}
where the equality holds if and only if $ q_{1}=lp_{1}t_{1}^{\frac{1}{\alpha}}$ and $q_{2}=lp_{2}t_{2}^{\frac{1}{\alpha}}$ with $l=\left(p_{1}t_{1}^{\frac{1}{\alpha}}+p_{2}t_{2}^{\frac{1}{\alpha}}\right)^{-1}$. Consequently
\begin{align*}
\mathop{\mathrm{max}}\limits_{q_{1},q_{2}}{(q_{1}^{1-\alpha}p_{1}^{\alpha}t_{1}+q_{2}^{1-\alpha}p_{2}^{\alpha}t_{2})}=\left(\sum_{m=1,2}p_{m}t_{m}^{\frac{1}{\alpha}}\right)^{\alpha}.
\end{align*}
Similarly, it is not difficult to obtain that when $\alpha>1$,
\begin{align*}
q_{1}^{1-\alpha}p_{1}^{\alpha}t_{1}+q_{2}^{1-\alpha}p_{2}^{\alpha}t_{2}\geq\left(\sum_{m=1,2}p_{m}t_{m}^{\frac{1}{\alpha}}\right)^{\alpha},
\end{align*}
and the equality holds when $q_{1}=lp_{1}t_{1}^{\frac{1}{\alpha}}$ and $q_{2}=lp_{2}t_{2}^{\frac{1}{\alpha}}$, which yields
\begin{align*}
\mathop{\mathrm{min}}\limits_{q_{1},q_{2}}{(q_{1}^{1-\alpha}p_{1}^{\alpha}t_{1}+q_{2}^{1-\alpha}p_{2}^{\alpha}t_{2})}=\left(\sum_{m=1,2}p_{m}t_{m}^{\frac{1}{\alpha}}\right)^{\alpha}.
\end{align*}
We have further
\begin{align*}
\Delta_{M_{\widetilde{\phi}}\in{\mathcal{I}}}f_{\alpha,z}^{\frac{1}{\alpha}}(M_{\phi},M_{\widetilde{\phi}})=
p_{1}\Delta_{M_{\widetilde{\phi}_{1}}\in{\mathcal{I}}}f_{\alpha,z}^{\frac{1}{\alpha}}
(M_{\phi_{1}},M_{\widetilde{\phi}_{1}})+
p_{2}\Delta_{M_{\widetilde{\phi}_{2}}\in{\mathcal{I}}}f_{\alpha,z}^{\frac{1}{\alpha}}
(M_{\phi_{2}},M_{\widetilde{\phi}_{2}}).
\end{align*}
Thus
\begin{align*}
&\mathit{C}_{\alpha,z}(\phi)=p_{1}\mathit{C}_{\alpha,z}(\phi_{1})+p_{2}\mathit{C}_{\alpha,z}(\phi_{2}),
\end{align*}
which implies that $\mathit{C}_{\alpha,z}(\phi)$ satisfies Eq. (\ref{eq4}). This completes the proof. $\hfill\qedsymbol$

\vskip0.1in

\noindent {\bf 3. An alternative coherence measure of quantum channels based on the generalized $\bm{\alpha}$-$\bm{z}$-relative R\'{e}nyi entropy}\\\hspace*{\fill}\\
In this section, we present a coherence measure of quantum channels through an alternative method by quantifying the commutativity between the channels and the completely dephasing channels via the generalized $\alpha$-$z$-relative R\'{e}nyi entropy. Furthermore, by utilizing the properties of the generalized $\alpha$-$z$-relative R\'{e}nyi entropy\cite{ZJF}, we discuss some properties of this coherence measure. \\\hspace*{\fill}\\
%$\textbf{Lemma    1}$ For arbitrary two quantum states $\rho$ and $\sigma$, $f_{\alpha,z}(\rho,\sigma)$ has the following properties\cite{ZJF}:\\
%$\left(1\right)$ $\alpha\in(0,1)$ and $z>0$ , $f_{\alpha,z}(\rho,\sigma)\leq{1}$; \\
%$\left(2\right)$ $\alpha\in(1,+\infty)$ and $z>0$ , $f_{\alpha,z}(\rho,\sigma)\geq{1}$;  \\ $\left(3\right)$ $f_{\alpha,z}(\rho,\sigma)=1\Leftrightarrow{\rho=\sigma}$.
%\\\hspace*{\fill}\\
%$\textbf{Lemma    2}$ For arbitrary two quantum states $\rho$ and $\sigma$, $D_{\alpha,z}(\rho,\sigma)$ has the following properties\cite{ZJF}:\\
%$\left(1\right)$ $D_{\alpha,z}\left(U\rho U^{\dagger},U\sigma U^{\dagger}\right)=D_{\alpha,z}(\rho,\sigma)$;\\
%$\left(2\right)$ For any quantum states $\rho$ and $\sigma$ with supp $\rho\subseteq$ supp $\sigma$, where supp $\rho $ and supp $\sigma$  represent the support of $\rho$ and $\sigma$ respectively. Suppose $\varepsilon$ is a CPTP map,
%for(i)-(iii)
%\begin{align*}
%D_{\alpha,z}\left(\varepsilon\left(\rho\right),\varepsilon\left(\sigma\right)\right)\leq{D_{\alpha,z}\left(\rho,\sigma\right)};
%\end{align*}
%(3) For (i)-(iii)
%\begin{align*}
%D_{\alpha,z}\left(\sum_{m}p_{m}\rho_{m},\sum_{m}p_{m}\sigma_{m}\right)\leq{\sum_{m}p_{m}D_{\alpha,z}(\rho_{m},\sigma_{m})}.
%\end{align*}
\noindent {\bf Definition 3}
The completely dephasing channel $\Delta^{A}\in{\mathcal{C}_{AB}}$ is defined as\cite{X}
\begin{equation}\label{eq9}
\Delta^{A}(\rho^{A})=\sum_{i}\langle
i|\rho^{A}|i\rangle|i\rangle\langle i|,
~~~\rho^{A}\in{\mathcal{D}(H_{A})}.
\end{equation}
A state $\sigma^{A}\in{\mathcal{D}(H_{A})}$ is called incoherent if $\Delta^{A}(\sigma^{A})=\sigma^{A}$. Otherwise we say that it is coherent.\\\hspace*{\fill}\\
%$\textbf{Definition    4}$  A channel $\phi\in{\mathcal{C}_{AB}}$ is an incoherent channel (IC)\cite{XJW} if
%$\gamma\left(\phi\right)=\phi,$ where $\gamma\left(\phi\right)=\Delta^{B}\phi\Delta^{A}$, i.e.,
%$\Delta^{B}\circ\phi\circ\Delta^{A}=\phi$.
%If a channel is not incoherent, we call that it is coherent.\\\hspace*{\fill}\\
\noindent {\bf Definition 4} For a channel $\phi\in{\mathcal{C}_{AB}}$, we define an alternative coherence measure $\widetilde{\mathit{C}}_{\alpha,z}(\phi)$ of $\phi$,
\begin{equation}\label{eq10}
\widetilde{\mathit{C}}_{\alpha,z}(\phi)=\sup_{\rho}D_{\alpha,z}
(\phi\circ\Delta^{A}(\rho),\Delta^{B}\circ\phi (\rho)),
\end{equation}
where $D_{\alpha,z}(\cdot , \cdot)$ is the generalized $\alpha$-$z$-relative R\'{e}nyi entropy, and the supremum in Eq. (\ref{eq10}) is taken over all quantum states.\\\hspace*{\fill}\\
\noindent {\bf Theorem 2} $\widetilde{\mathit{C}}_{\alpha,z}(\phi)$ has the following elegant properties:\\
(i) (extremal property) for $\sup\limits_{\rho}D_{\alpha,z}(\phi\circ\Delta(\rho),\Delta\circ\phi(\rho))$, there exits a pure state $|\psi\rangle\langle\psi|$ such that the supremum in Eq. (\ref{eq10}) is attained when $\rho=|\psi\rangle\langle\psi|$.\\
(ii) (monotonicity) for any quantum channel $\phi$, if $\phi_{0}$ is a quantum channel satisfying $\widetilde{\mathit{C}}_{\alpha,z}(\phi_{0})=0$, then $\widetilde{\mathit{C}}_{\alpha,z}(\phi_{0}\circ\phi)\leq \widetilde{\mathit{C}}_{\alpha,z}(\phi)$  and $\widetilde{\mathit{C}}_{\alpha,z}(\phi\circ\phi_{0})\leq \widetilde{\mathit{C}}_{\alpha,z}(\phi)$.\\
(iii) (convexity) for some quantum channels $\phi_{m}$, and some positive real number $\lambda_{m}$ such that $\sum\limits_{m}\lambda_{m}=1$, we have
$\widetilde{\mathit{C}}_{\alpha,z}\left(\sum\limits_{m}\lambda_{m}\phi_{m}\right)
\leq\sum\limits_{m}\lambda_{m}\widetilde{\mathit{C}}_{\alpha,z}
\left(\phi_{m}\right)$.\\\hspace*{\fill}\\
%We give the concrete proof as follows.\\\hspace*{\fill}\\
{\bf Proof} Suppose that the spectral decomposition of $\rho$ is
$\rho=\sum\limits_{m}\mu_{m}|\psi_{m}\rangle\langle\psi_{m}|$. We
have
\begin{align*}
&D_{\alpha,z}(\phi\circ\Delta(\rho),\Delta\circ\phi(\rho))\\
=&D_{\alpha,z}\left(\phi\circ\Delta\left(\sum_{m}\mu_{m}| \psi_{m}\rangle\langle\psi_{m}|\right),\Delta\circ\phi\left(\sum_{m}\mu_{m}|\psi_{m}\rangle\langle\psi_{m}|\right)\right)\\
=&D_{\alpha,z}\left(\sum_{m}\mu_{m}\phi\circ\Delta(|\psi_{m}\rangle\langle\psi_{m}|),\sum_{m}\mu_{m}\Delta\circ\phi(|\psi_{m}\rangle\langle\psi_{m}|)\right)\\
\leq&\sum_{m}\mu_{m}D_{\alpha,z}(\phi\circ\Delta(|\psi_{m}\rangle\langle\psi_{m}|),\Delta\circ\phi(|\psi_{m}\rangle\langle\psi_{m}|))\\
\leq&\sum_{m}\mu_{m}\sup_{|\psi\rangle}D_{\alpha,z}(\phi\circ\Delta(|\psi\rangle\langle\psi|),\Delta\circ\phi(|\psi\rangle\langle\psi|))\\
=&\sup_{|\psi\rangle}D_{\alpha,z}(\phi\circ\Delta(|\psi\rangle\langle\psi|),\Delta\circ\phi(|\psi\rangle\langle\psi|)),
\end{align*}
where the first inequality follows from the joint convexity of $D_{\alpha,z}(\cdot,\cdot)$.
Thus,
\begin{align*}
%&D_{\alpha,z}(\phi\circ\Delta(\rho),\Delta\circ\phi(\rho))\leq\sup_{|\psi\rangle}D_{\alpha,z}(\phi\circ\Delta(|\psi\rangle\langle\psi|),\Delta\circ\phi(|\psi\rangle\langle\psi|)),\\
&\widetilde{\mathit{C}}_{\alpha,z}(\phi)\leq\sup_{|\psi\rangle}D_{\alpha,z}(\phi\circ\Delta(|\psi\rangle\langle\psi|),\Delta\circ\phi(|\psi\rangle\langle\psi|)).
\end{align*}
It follows from Eq. (\ref{eq10}) that
\begin{equation}\label{eq11}
\widetilde{\mathit{C}}_{\alpha,z}(\phi)=\sup_{|\psi\rangle}D_{\alpha,z}(\phi\circ\Delta(|\psi\rangle\langle\psi|),\Delta\circ\phi(|\psi\rangle\langle\psi|)).
\end{equation}
Therefore, item (i) holds.

Using the monotonicity of $D_{\alpha,z}$ under the CPTP maps, we have
\begin{align*}
&D_{\alpha,z}(\phi_{0}\circ\phi\circ\Delta(|\psi\rangle\langle\psi|),\Delta\circ\phi_{0}\circ\phi(|\psi\rangle\langle\psi|))\\
=&D_{\alpha,z}(\phi_{0}\circ\phi\circ\Delta(|\psi\rangle\langle\psi|),\phi_{0}\circ\Delta\circ\phi(|\psi\rangle\langle\psi|))\\
\leq& D_{\alpha,z}(\phi\circ\Delta(|\psi\rangle\langle\psi|),\Delta\circ\phi(|\psi\rangle\langle\psi|)),
\end{align*}
where the first equality holds due to $\widetilde{\mathit{C}}_{\alpha,z}(\phi_{0})=0$ and Definition $1$ in\cite{FGY}. Then by Eq. (\ref{eq11}), we obtain
$\widetilde{\mathit{C}}_{\alpha,z}(\phi_{0}\circ\phi)\leq \widetilde{\mathit{C}}_{\alpha,z}(\phi)$. On the other hand,
\begin{align*}
&\widetilde{\mathit{C}}_{\alpha,z}(\phi\circ\phi_{0})\\
=&\sup_{\rho}D_{\alpha,z}(\phi\circ\phi_{0}\circ\Delta(\rho),\Delta\circ\phi\circ\phi_{0}(\rho))\\
=&\sup_{\rho}D_{\alpha,z}(\phi\circ\Delta\circ\phi_{0}(\rho),\Delta\circ\phi\circ\phi_{0}(\rho))\\
=&\sup_{\sigma=\phi_{0}(\rho)}D_{\alpha,z}(\phi\circ\Delta(\sigma),\Delta\circ\phi(\sigma))\\
\leq&\sup_{\rho}D_{\alpha,z}(\phi\circ\Delta(\rho),\Delta\circ\phi(\rho))\\
=&\widetilde{\mathit{C}}_{\alpha,z}(\phi),
\end{align*}
which implies that $\widetilde{\mathit{C}}_{\alpha,z}(\phi\circ\phi_{0})\leq \widetilde{\mathit{C}}_{\alpha,z}(\phi)$. Hence, item (ii) is proved.

By utilizing the joint convexity of $D_{\alpha,z}(\cdot,\cdot)$, we can further obtain
\begin{align*}
&\widetilde{\mathit{C}}_{\alpha,z}\left(\sum_{m}\lambda_{m}\phi_{m}\right)\\
=&\sup_{|\psi\rangle} D_{\alpha,z}\left(\sum_{m}\lambda_{m}\phi_{m}\circ\Delta(|\psi\rangle\langle\psi|),\Delta\circ\sum_{m}\lambda_{m}\phi_{m}(|\psi\rangle\langle\psi|)\right)\\
=&\sup_{|\psi\rangle} D_{\alpha,z}\left(\sum_{m}\lambda_{m}\phi_{m}\circ\Delta(|\psi\rangle\langle\psi|),\sum_{m}\lambda_{m}\Delta\circ\phi_{m}(|\psi\rangle\langle\psi|)\right)\\
\leq&\sum_{m}\lambda_{m}\sup_{|\psi\rangle}D_{\alpha,z}\left(\phi_{m}\circ\Delta(|\psi\rangle\langle\psi|),\Delta\circ\phi_{m}(|\psi\rangle\langle\psi|)\right)\\
=&\sum_{m}\lambda_{m}\widetilde{\mathit{C}}_{\alpha,z}(\phi_{m}).
\end{align*}
Therefore
\begin{equation}\label{eq12}
\widetilde{\mathit{C}}_{\alpha,z}\left(\sum_{m}\lambda_{m}\phi_{m}\right)\leq\sum_{m}\lambda_{m}\widetilde{\mathit{C}}_{\alpha,z}(\phi_{m}),
\end{equation}
and the item (iii) is derived. $\hfill\qedsymbol$

From Eq. (\ref{eq10}), it can be easily seen that
$\widetilde{\mathit{C}}_{\alpha,z}(\phi)=0$ when the quantum channel
$\phi$ is detection-creation-incoherent\cite{FGY}, i.e.,
$\phi\circ\Delta^{A}=\Delta^{B}\circ\phi$. Comparing the two
quantifiers of the coherence of quantum channels in Eqs. (\ref{eq8})
and (\ref{eq10}), it can be found that $\mathit{C}_{\alpha,z}(\phi)
\geq \widetilde{\mathit{C}}_{\alpha,z}(\phi)$ always holds in this
special case. From the examples in the next section and numerical
results, it is conjectured that $\mathit{C}_{\alpha,1}(\phi) \geq
\widetilde{\mathit{C}}_{\alpha,1}(\phi)$ holds for all quantum
channels $\phi$, but we have not yet found a proof.

\vskip0.1in

\noindent {\bf 4. Examples}\\\hspace*{\fill}\\
In this section, we choose several typical channels to calculate the
coherence measures defined in Eqs. (\ref{eq8}) and
(\ref{eq10}).$\\\hspace*{\fill}\\$ \noindent {\bf  Example 1}
Consider the phase flip channel $\phi_{PF}(\rho)
=\sum\limits_{n=1}^{2}K_{n}\rho K_{n}^{\dagger}$ with the Kraus
operators
\begin{align*}
&K_{1}=\sqrt{p}\left(\begin{array}{cc} 1 & 0\\0 & 1\\\end{array}\right),~~ K_{2}=\sqrt{1-p}\left(\begin{array}{cc} 1 & 0\\0 & -1\\\end{array}\right),~~~0\leq p\leq1.
\end{align*}
Direct calculation shows that
 \begin{align}\label{eq13}
 \mathit{C}_{\alpha,1}(\phi_{PF})&=\frac{\sum\limits_{i,\beta=0}^{1}\langle i\beta|\mathit{M}_{\phi_{PF}}^{\alpha}|i\beta\rangle^{\frac{1}{\alpha}}-1}{\alpha-1}
 =\frac{2^{1-\frac{1}{\alpha}}[p^{\alpha}+(1-p)^{\alpha}]^{\frac{1}{\alpha}}-1}{\alpha-1}.
 \end{align}

However, if we calculate the values of the coherence measure given in Eq. (\ref{eq10}), we can clearly see that $\widetilde{\mathit{C}}_{\alpha,z}\left(\phi_{PF}\right)\equiv0$ regardless of the values of $\alpha$ and $z$. In fact, for any pure state
$|\psi\rangle=a|0\rangle+b|1\rangle$ with $|a|^{2}+|b|^{2}=1$, we have
\begin{align*}
\Delta\circ\phi_{PF}(|\psi\rangle\langle\psi|)
&=\Delta(\phi_{PF}(|\psi\rangle\langle\psi|))\\
&=\Delta(K_{1}(|\psi\rangle\langle\psi|)K_{1}^{\dagger} +K_{2}(|\psi\rangle\langle\psi|)K_{2}^{\dagger}),
\end{align*}
where $K_{1}|\psi\rangle=a\sqrt{p}|0\rangle+b\sqrt{p}|1\rangle$ and
$K_{2}|\psi\rangle=a\sqrt{1-p}|0\rangle-b\sqrt{1-p}|1\rangle$.
It can be shown that
\begin{align*}
&\phi_{PF}(|\psi\rangle\langle\psi|)=|a|^{2}|0\rangle\langle0|+(2p-1)a\bar{b}|0\rangle\langle1|+(2p-1)b\bar{a}|1\rangle\langle0|+|b|^{2}|1\rangle\langle1|,\\
&\Delta\circ\phi_{PF}(|\psi\rangle\langle\psi|)=|a|^{2}|0\rangle\langle0|+|b|^{2}|1\rangle\langle1|,\\
&\phi_{PF}\circ\Delta(|\psi\rangle\langle\psi|)=\phi_{PF}(|a|^{2}|0\rangle\langle0|+|b|^{2}|1\rangle\langle1|)=|a|^{2}|0\rangle\langle0|+|b|^{2}|1\rangle\langle1|,
\end{align*}
which implies that
$\widetilde{\mathit{C}}_{\alpha,z}(\phi_{PF})=0$.

\begin{figure}[H]\centering
{\begin{minipage}[Figure-1]{0.5\linewidth}
\includegraphics[width=0.95\textwidth]{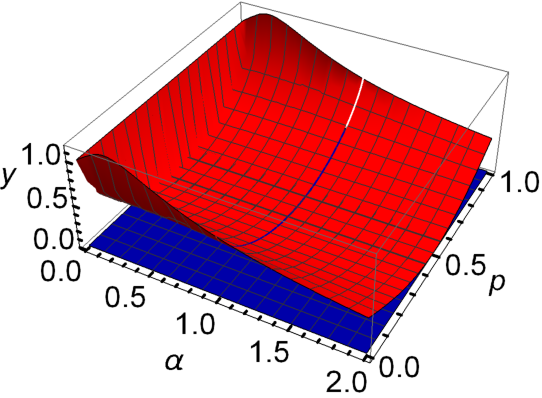}
\end{minipage}}
\caption{{Surfaces of $\widetilde{\mathit{C}}_{\alpha,z}(\phi_{PF})$ and $\mathit{C}_{\alpha,1}(\phi_{PF})$. The blue (red) surface represents the values of $\widetilde{\mathit{C}}_{\alpha,z}(\phi_{PF})$ $(\mathit{C}_{\alpha,1}(\phi_{PF}))$
in Eq. (\ref{eq10}) (Eq. (\ref{eq13})).\label{fig:Fig1}}}
\end{figure}

In Fig.~\ref{fig:Fig1}, we plot the surfaces of
$\widetilde{\mathit{C}}_{\alpha,z}(\phi_{PF})$ and
$\mathit{C}_{\alpha,1}(\phi_{PF})$ given in Eqs. (\ref{eq10}) and
(\ref{eq13}). By calculation, it is found that
$\lim\limits_{\alpha\rightarrow1}\mathit{C}_{\alpha,1}(\phi_{PF})=\mathrm{ln2}+p\mathrm{lnp}+\mathrm{ln}\left(1-p\right)-p\mathrm{ln}\left(1-p\right)$,
which reaches its minimum value $0$ when $p=\frac{1}{2}$, and
reaches its maximum value $\mathrm{ln2}$ when $p=0$. When
$\alpha=\frac{1}{2}$, $\mathit{C}_{\frac{1}{2},1}(\phi_{PF})=1-2\sqrt{p(1-p)}$. Its
minimum value $0$ is obtained when $p=\frac{1}{2}$ and its
maximum value $1$ is obtained when $p=0$. It can be shown that
$\mathit{C}_{\alpha,1}(\phi_{PF}) \geq
\widetilde{\mathit{C}}_{\alpha,z}(\phi_{PF})$ when
$\alpha\in(0,1)\cup(1,2]$, $0\leq p\leq1$.

\noindent {\bf  Example 2} Consider the depolarizing channel
$\phi_{D}(\rho)=\sum\limits_{n=1}^{4}K_{n}\rho K_{n}^{\dagger}$ with
the Kraus operators
\begin{align*}
&K_{1}=\sqrt{1-\frac{3}{4}p}\left(\begin{array}{cc} 1 & 0\\0 & 1\\\end{array}\right), ~~~K_{2}=\frac{\sqrt{p}}{2}\left(\begin{array}{cc} 0 & 1\\1 & 0\\\end{array}\right),\\ &K_{3}=\frac{\sqrt{p}}{2}\left(\begin{array}{cc} 0 & -\mathrm{i}\\\mathrm{i} & 0\\\end{array}\right),
~~~~~~~~K_{4}=\frac{\sqrt{p}}{2}\left(\begin{array}{cc} 1 & 0\\0 & -1\\\end{array}\right),~~~0\leq p\leq1.
\end{align*}
Hence $\mathit{C}_{\alpha,1}(\phi)$ defined in Eq. (\ref{eq8}) is given by
\begin{align}\label{eq14}
\mathit{C}_{\alpha,1}(\phi_{D})&=\frac{\sum\limits_{i,\beta=0}^{1}\langle i\beta|\mathit{M}_{\phi_{D}}^{\alpha}|i\beta\rangle
^{\frac{1}{\alpha}}-1}{\alpha-1}=\frac{2\left[\frac{p^{\alpha}}
{2^{2\alpha+1}}+\frac{\left(1-\frac{3}{4}p\right)^{\alpha}}
{2}\right]^{\frac{1}{\alpha}}+\frac{p}{2}-1}{\alpha-1}.
\end{align}
Similar to the phase flip channel, $\widetilde{\mathit{C}}_{\alpha,z}\left(\phi_{D}\right)\equiv0$ regardless of the values of $\alpha$ and $z$.

\begin{figure}[H]\centering
{\begin{minipage}[Figure-2]{0.5\linewidth}
\includegraphics[width=0.95\textwidth]{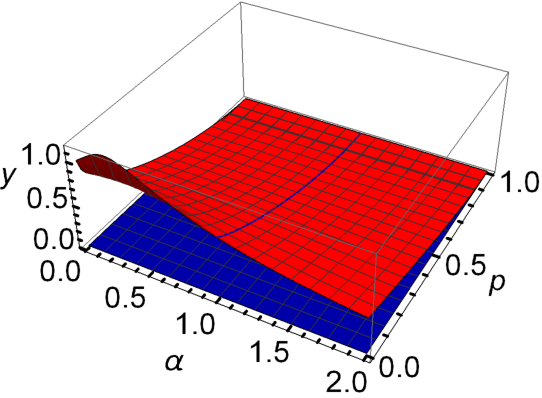}
\end{minipage}}
\caption{{Surfaces of $\widetilde{\mathit{C}}_{\alpha,z}(\phi_{D})$ and $\mathit{C}_{\alpha,1}(\phi_{D})$. The blue (red) surface represents the values of $\widetilde{\mathit{C}}_{\alpha,z}(\phi_{D})$ $(\mathit{C}_{\alpha,1}(\phi_{D}))$
in Eq. (\ref{eq10}) (Eq. (\ref{eq14})).\label{fig:Fig2}}}
\end{figure}

In Fig.~\ref{fig:Fig2}, we plot the surfaces of
$\widetilde{\mathit{C}}_{\alpha,z}(\phi_{D})$ and
$\mathit{C}_{\alpha,1}(\phi_{D})$ in Eqs. (\ref{eq10}) and
(\ref{eq14}). Direct calculation shows that
$\lim\limits_{\alpha\rightarrow1}\mathit{C}_{\alpha,1}(\phi_{D})=\frac{1}{4}[(4-3p)\mathrm{ln}(4-3p)+
2(p-2)\mathrm{ln}(2-p)+p\mathrm{ln}p]$, which reaches its minimum
value $0$ when $p=1$, and reaches its maximum value
$\mathrm{ln2}$ when $p=0$. When $\alpha=\frac{1}{2}$, we have
$\mathit{C}_{\frac{1}{2},1}(\phi_{D})=1-\frac{\sqrt{p(4-3p)}+p}{2}$.
Its minimum value $0$ is attained when $p=1$, and its maximum
value of $1$ is attained when $p=0$. It can be found that
$\mathit{C}_{\alpha,1}(\phi_{D})\geq\widetilde{\mathit{C}}_{\alpha,z}(\phi_{D})$
when $\alpha\in(0,1)\cup(1,2]$, $0\leq p\leq1$.

\noindent {\bf  Example 3} Consider the amplitude damping channel
$\phi_{AD}(\rho)=\sum\limits_{n=1}^{2}K_{n}\rho K_{n}^{\dagger}$
with the Kraus operators
\begin{align*}
&K_{1}=\left(\begin{array}{cc} 1 & 0\\0 & \sqrt{1-p}\\\end{array}\right),~~ K_{2}=\left(\begin{array} {cc} 0 & \sqrt{p}\\0 & 0\\\end{array}\right),~~~0\leq p\leq1.
\end{align*}
It follows from Eq. (\ref{eq8}) that
\begin{align}\label{eq15}
\mathit{C}_{\alpha,1}(\phi_{AD})&=\frac{\sum\limits_{i,\beta=0}^{1}\langle i\beta|\mathit{M}_{\phi_{AD}}^{\alpha}|i\beta\rangle^{\frac{1}{\alpha}}-1}{\alpha-1}=\frac{\left(\frac{1}{2}+\frac{1}{2}\left(1-p\right)^{\frac{1}{\alpha}}\right)(2-p)^{1-\frac{1}{\alpha}}+\frac{p}{2}-1}{\alpha-1}.
\end{align}
Similarly, $\widetilde{\mathit{C}}_{\alpha,z}\left(\phi_{AD}\right)\equiv0$ regardless of the values of $\alpha$ and $z$.

\begin{figure}[H]\centering
{\begin{minipage}[Figure-3]{0.5\linewidth}
\includegraphics[width=0.95\textwidth]{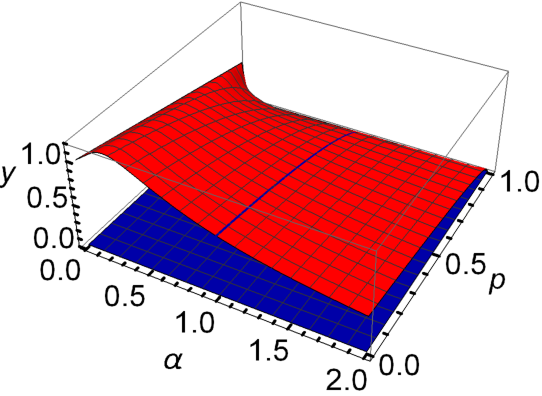}
\end{minipage}}
\caption{{Surfaces of $\widetilde{\mathit{C}}_{\alpha,z}(\phi_{AD})$ and $\mathit{C}_{\alpha,1}(\phi_{AD})$. The blue (red) surface represents the values of $\widetilde{\mathit{C}}_{\alpha,z}(\phi_{AD})$ $(\mathit{C}_{\alpha,1}(\phi_{AD}))$
in Eq. (\ref{eq10}) (Eq. (\ref{eq15})).\label{fig:Fig3}}}
\end{figure}

In Fig.~\ref{fig:Fig3}, we plot the surfaces of
$\widetilde{\mathit{C}}_{\alpha,z}(\phi_{AD})$ and
$\mathit{C}_{\alpha,1}(\phi_{AD})$ in Eqs. (\ref{eq10}) and
(\ref{eq15}). It is found that
$\lim\limits_{\alpha\rightarrow1}\mathit{C}_{\alpha,1}(\phi_{AD})
=\frac{1}{2}[(p-1)\mathrm{ln}(1-p)-(p-2)\mathrm{ln}(2-p)]$.
$\lim\limits_{\alpha\rightarrow1}\mathit{C}_{\alpha,1}(\phi_{AD})$
reaches its minimum value $0$ when $p=1$.
$\lim\limits_{\alpha\rightarrow1}\mathit{C}_{\alpha,1}(\phi_{AD})$
reaches its maximum value $\mathrm{ln2}$ when $p=0$. When
$\alpha=\frac{1}{2}$, we have
$\mathit{C}_{\frac{1}{2},1}(\phi_{AD})=\frac{2p-2}{p-2}$. Its
minimum value $0$ is obtained when $p=1$ and its maximum value
$1$ is obtained when $p=0$.  It can be shown that
$\mathit{C}_{\alpha,1}(\phi_{AD}) \geq
\widetilde{\mathit{C}}_{\alpha,z}(\phi_{AD})$ when
$\alpha\in(0,1)\cup(1,2]$, $0\leq p\leq1$.

\noindent {\bf  Example 4} Consider the isotropic channel
$\phi_{\Lambda}$ for $t\in[\frac{-1}{d^{2}-1},1]$\cite{LN}
\begin{align}\label{eq16}
\phi_{\Lambda}(\rho)=tU \rho
U^{\dagger}+(1-t)\frac{\mathbb{I}_d}{d},
\end{align}
where $U$ is an unitary operation, $\mathbb{I}_d$ is $d\times d$
identity matrix, and $d$ is the dimension of the Hilbert space. In
particular, taking $U=H$, where
$H=\frac{1}{\sqrt{2}}\left(\begin{array}{cc} 1 & 1\\1 &
-1\\\end{array}\right)$ is the Hadamard gate, we have
\begin{align}\label{eq17}
\phi_{\Lambda}^{H}(\rho)=tH \rho
H^{\dagger}+(1-t)\frac{\mathbb{I}_2}{2}
=\sum\limits_{n=1}^{5}K_{n}\rho K_{n}^{\dagger},~~~-\frac{1}{3}\leq
t\leq1,
\end{align}
where $\mathbb{I}_2$ is $2\times 2$ identity matrix, and
\begin{align*}
&K_{1}=\sqrt{t}H=\sqrt{\frac{t}{2}}\left(\begin{array}{cc} 1 & 1\\1
& -1\\\end{array}\right),
~~~~~~~~~~~~~K_{2}=\frac{\sqrt{1-t}}{2}X=\frac{\sqrt{1-t}}{2}\left(\begin{array}{cc}
0 & 1\\1 & 0\\\end{array}\right),\\
&K_{3}=\frac{\sqrt{1-t}}{2}Y=\frac{\sqrt{1-t}}{2}\left(\begin{array}{cc}
0 & -\mathrm{i}\\\mathrm{i} & 0\\\end{array}\right),
~~~~K_{4}=\frac{\sqrt{1-t}}{2}Z=\frac{\sqrt{1-t}}{2}\left(\begin{array}{cc}
1 & 0\\0 & -1\\\end{array}\right), \\
&K_{5}=\frac{\sqrt{1-t}}{2}\mathbb{I}_2=\frac{\sqrt{1-t}}{2}\left(\begin{array}{cc}
1 & 0\\0 & 1\\\end{array}\right).
\end{align*}
By Eq.(\ref{eq8}), it can be easily deduced that
\begin{align}\label{eq18}
\mathit{C}_{\alpha,1}({\phi_{\Lambda}^{H}})
&=\frac{\sum\limits_{i,\beta=0}^{1}\langle
i\beta|\mathit{M}_{\phi_{\Lambda}^{H}}^{\alpha}|i\beta\rangle^{\frac{1}{\alpha}}-1}{\alpha-1}
=\frac{4^{-\frac{1}{\alpha}}[3(1-t)^{\alpha}+(1+3t)^{\alpha}]^{\frac{1}{\alpha}}-1}{\alpha-1}.
\end{align}
According to Eq. (\ref{eq18}), we obtain
\begin{align}\label{eq19}
\lim\limits_{\alpha\rightarrow1}\mathit{C}_{\alpha,1}(\phi_{\Lambda}^{H})=
\frac{3(1-t)\mathrm{ln}(1-t)+(1+3t)\mathrm{ln}(1+3t)}{4},
\end{align}
\begin{align}\label{eq20}
\mathit{C}_{\frac{1}{2},1}(\phi_{\Lambda}^{H})=\frac{3t-5}{4}-\frac{3}{4}\sqrt{(1-t)(1-3t)}+2.
\end{align}
%Noting that
%\begin{align*}
%\mathit{M}_{\phi_{H}}
%=&(\mathbb{I}\otimes H)\left(\frac{1}{2}\sum_{i,j=0}^{1}|ii\rangle\langle jj|\right)(\mathbb{I}\otimes H)^{\dagger}\\
%=&\frac{1}{4}\left(\begin{array}{cccc} 1 & 1 & 1 & -1\\ 1 & 1 & 1 &
%-1\\ 1 & 1 & 1 & -1\\-1 & -1 & -1 & 1\\\end{array}\right),
%\end{align*}
%we have
%\begin{align*}
%\mathit{M}_{\phi_{H}}^{\alpha}
%=&\frac{1}{4}\left(\begin{array}{cccc}
%1 & 1 & 1 & -1\\1 & 1 & 1 & -1\\1 & 1 & 1 & -1\\ -1 & -1 & -1 & 1\\\end{array}\right).
%\end{align*}
Set $\alpha=\frac{1}{2}$ and $z=1$. Then
\begin{align*}
&\widetilde{\mathit{C}}_{\frac{1}{2},1}(\phi_{\Lambda}^{H})
=\sup_{|\psi\rangle}D_{\frac{1}{2},1}(\phi_{\Lambda}^{H}\circ\Delta(|\psi\rangle|\langle\psi|),\Delta\circ
\phi_{\Lambda}^{H}(|\psi\rangle|\langle\psi|)),
\end{align*}
where
\begin{align*}
&D_{\frac{1}{2},1}(\phi_{\Lambda}^{H}\circ\Delta(|\psi\rangle|\langle\psi|),\Delta\circ
\phi_{\Lambda}^{H}(|\psi\rangle|\langle\psi|))\\
=&-2[f_{\frac{1}{2},1}^{2}(\phi_{\Lambda}^{H}\circ\Delta(|\psi\rangle|\langle\psi|),\Delta\circ \phi_{\Lambda}^{H}(|\psi\rangle|\langle\psi|))-1]\\
=&-2\left[\left[\mathrm{Tr}\left((\phi_{\Lambda}^{H}\circ\Delta(|\psi\rangle|\langle\psi|))^{\frac{1}{2}}(\Delta\circ \phi_{\Lambda}^{H}(|\psi\rangle|\langle\psi|))^{\frac{1}{2}}\right)\right]^{2}-1\right]\\
=&-2\left[\left(1+\sqrt{1-4t^{2}\mathrm{Re}^{2}(ab^{\ast})}\right)\left(\frac{1}{4}+\frac{1}{4}
\sqrt{1-t^{2}(
\vert{a}\vert^{2}-\vert{b}\vert^{2})^{2}}\right)-1\right]\\
\leq&-2\left[\left(1+\sqrt{1-4t^{2}\vert{a}\vert^{2}\vert{b}\vert^{2}}\right)\left(\frac{1}{4}+\frac{1}{4}
\sqrt{1-t^{2}(
\vert{a}\vert^{2}-\vert{b}\vert^{2})^{2}}\right)-1\right]\\
\leq&-2\left[\left(1+\sqrt{1-t^{2}}\right)\left(\frac{1}{4}+\frac{1}{4}
\sqrt{1-t^{2}(
\vert{a}\vert^{2}-\vert{b}\vert^{2})^{2}}\right)-1\right]\\
\leq&-2\left[\left(1+\sqrt{1-t^{2}}\right)\left(\frac{1}{4}+\frac{1}{2}
\vert{a}\vert\vert{b}\vert\right)-1\right]\\
\leq&-2\vert{a}\vert\vert{b}\vert\left(1+\sqrt{1-t^{2}}\right)+2\\
\leq&1-\sqrt{1-t^{2}}.
\end{align*}
The above inequalities hold due to the facts that
$0\leq\vert{a}\vert^{2}\vert{b}\vert^{2}\leq \frac{1}{4}$ and $(
\vert{a}\vert^{2}-\vert{b}\vert^{2})^{2}=1-4\vert{a}\vert^{2}\vert{b}\vert^{2}$.
It follows from item (i) that
$\widetilde{\mathit{C}}_{\frac{1}{2},1}(\phi_{\Lambda}^{H})\leq1-\sqrt{1-t^{2}}$.
Meanwhile, for the classical pure state $|0\rangle$ or $|1\rangle$,
the maximum value of $D_{\frac{1}{2},1}(\Delta\circ
\phi_{\Lambda}^{H}(\rho),\phi_{\Lambda}^{H}\circ\Delta(\rho))$ can
be obtained directly. It is easy to see that
\begin{align*}
D_{\frac{1}{2},1}(\Delta\circ
\phi_{\Lambda}^{H}(|0\rangle\langle0|),\phi_{\Lambda}^{H}\circ\Delta(|0\rangle\langle0|))=1-\sqrt{1-t^{2}}.
\end{align*}
Thus we get
\begin{equation}\label{eq21}
\widetilde{\mathit{C}}_{\frac{1}{2},1}(\phi_{\Lambda}^{H})=1-\sqrt{1-t^{2}}.
\end{equation}
According to the above results, it is found that
$\widetilde{\mathit{C}}_{\alpha,z}(\phi_{\Lambda}^{H})$ is not an
incoherent channel when $\alpha=\frac{1}{2}$ and $z=1$.

Setting $t=1$ in Eq. (\ref{eq17}), $\phi_{\Lambda}^H$ becomes the
unitary channel $\phi_{H}$ induced by the Hadamard gate $H$. Then it
follows from Eq. (\ref{eq18}) that
\begin{align}\label{eq22}
\mathit{C}_{\alpha,1}({\phi_{H}})&=\frac{\sum\limits_{i,\beta=0}^{1}\langle i\beta|\mathit{M}_{{\phi_{H}}}^{\alpha}|i\beta\rangle^{\frac{1}{\alpha}}-1}{\alpha-1}
=\frac{4^{1-\frac{1}{\alpha}}-1}{\alpha-1}.
\end{align}
According to Eq. (\ref{eq22}), we obtain that
$\lim\limits_{\alpha\rightarrow1}\mathit{C}_{\alpha,1}(\phi_{H})=\mathrm{ln4}$
and $\mathit{C}_{\frac{1}{2},1}(\phi_{H})=\frac{3}{2}$. From the
deduction of
$\widetilde{\mathit{C}}_{\frac{1}{2},1}(\phi_{\Lambda}^H)$, we can
also infer that $\widetilde{\mathit{C}}_{\frac{1}{2},1}(\phi_{H})=1$
by letting $t=1$.

It can be seen that
$\mathit{C}_{\frac{1}{2},1}(\phi_{\Lambda}^H)\geq
\widetilde{\mathit{C}}_{\frac{1}{2},1}(\phi_{\Lambda}^H)$ holds when
$-\frac{1}{3}\leq t\leq1$. And as a special case of $t=1$, we get
$\mathit{C}_{\frac{1}{2},1}(\phi_{H})\geq\widetilde{\mathit{C}}_{\frac{1}{2},1}(\phi_{H})$.
In Fig. 4, we plot the values of
$\mathit{C}_{\frac{1}{2},1}(\phi_{\Lambda}^H)$ and $
\widetilde{\mathit{C}}_{\frac{1}{2},1}(\phi_{\Lambda}^H)$ in Eqs.
(\ref{eq20}) and (\ref{eq21}).
\begin{figure}[H]\centering
{\begin{minipage}[Figure-4]{0.5\linewidth}
\includegraphics[width=1.5\textwidth]{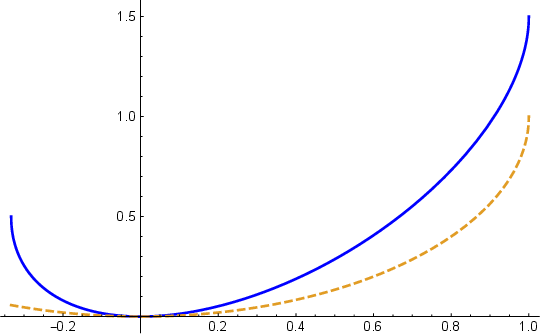}
\end{minipage}}
\caption{{The values of
$\mathit{C}_{\frac{1}{2},1}(\phi_{\Lambda}^H)$ and
$\widetilde{\mathit{C}}_{\frac{1}{2},1}(\phi_{\Lambda}^H)$. The blue
(orange) curve represents the values of
$\mathit{C}_{\frac{1}{2},1}(\phi_{\Lambda}^H)$
($\widetilde{\mathit{C}}_{\frac{1}{2},1}(\phi_{\Lambda}^H)$) in Eq.
(\ref{eq20}) (Eq. (\ref{eq21})).\label{fig:Fig4}}}
\end{figure}
\vskip0.1in

\noindent {\bf  Example 5} Consider the unitary channels $\phi_{S}$
and $\phi_{T}$ induced by the phase gate $S$ and $\frac{\pi}{8}$
gate $T$, i.e., $\phi_{S}(\rho)=S \rho S^{\dagger}$ and
$\phi_{T}(\rho)=T \rho T ^{\dagger}$, where
\begin{align*}
S=\left(\begin{array}{cc} 1 & 0\\0 &
\mathrm{i}\\\end{array}\right)~~~\makebox{and}~~~
T=\left(\begin{array}{cc} 1 & 0\\0 &
e^{\frac{\mathrm{i}\pi}{4}}\\\end{array}\right).
\end{align*}
By Eq. (\ref{eq8}), we have $\mathit{C}_{\alpha,1}(\phi_{S})=
\mathit{C}_{\alpha,1}(\phi_{T})=\frac{2^{1-\frac{1}{\alpha}}-1}{\alpha-1}$.
It is obvious that
$\lim\limits_{\alpha\rightarrow1}\mathit{C}_{\alpha,1}(\phi_{S})
=\lim\limits_{\alpha\rightarrow1}\mathit{C}_{\alpha,1}(\phi_{T})=
\mathrm{ln2}$, and
$\mathit{C}_{\frac{1}{2},1}(\phi_{S})=\mathit{C}_{\frac{1}{2},1}(\phi_{T})=1$.
By Eq. (\ref{eq10}) we obtain
$\widetilde{\mathit{C}}_{\alpha,z}(\phi_{S})=
\widetilde{\mathit{C}}_{\alpha,z}(\phi_{T})=0$. Note that the two
quantifiers of the coherence $\mathit{C}_{\alpha,1}(\cdot)$ and
$\widetilde{\mathit{C}}_{\alpha,1}(\cdot)$ for the quantum channels
induced by $S$ and $T$ are the same.

From Examples 4 and 5, it can be seen that
$\mathit{C}_{\frac{1}{2},1}(\phi)>
\widetilde{\mathit{C}}_{\frac{1}{2},1}(\phi)$, where $\phi$ is the
unitary channel induced by $H$, $S$ or $T$.

The above results are based on the channels of single qubits. We now
turn to discuss the channels of entangled qubits. The corresponding
Choi-Jamio{\l}kowski states for the channels of entangled qubits are
too complicated to be calculated for general two-qubit unitaries.
For simplicity, we take $S\otimes S$ and $T\otimes T$.

\noindent {\bf  Example 6} Consider the unitary channels
$\phi_{S\otimes S}$ and $\phi_{T\otimes T}$ induced by $S\otimes S$
and $T\otimes T$, i.e., $\phi_{S\otimes S}(\rho_{AB})=(S\otimes S)
\rho_{AB} (S\otimes S)^{\dagger}$ and $\phi_{T\otimes
T}(\rho_{AB})=(T\otimes T) \rho_{AB} (T \otimes T)^{\dagger}$, where
$S$ is the phase gate and $T$
is the $\frac{\pi}{8}$ gate defined in Example 5.\\
By Eq. (\ref{eq8}), it follows that
\begin{align}\label{eq23}
\mathit{C}_{\alpha,1}(\phi_{S\otimes
S})=\mathit{C}_{\alpha,1}(\phi_{T\otimes
T})=\frac{4^{1-\frac{1}{\alpha}}-1}{\alpha-1}.
\end{align}
It is obvious that
$\lim\limits_{\alpha\rightarrow1}\mathit{C}_{\alpha,1}(\phi_{S\otimes
S})
=\lim\limits_{\alpha\rightarrow1}\mathit{C}_{\alpha,1}(\phi_{T\otimes
T})= \mathrm{ln4}$, and $\mathit{C}_{\frac{1}{2},1}(\phi_{S\otimes
S})=\mathit{C}_{\frac{1}{2},1}(\phi_{T\otimes T})=\frac{3}{2}$. On
the other hand, by using Eq. (\ref{eq10}) we obtain $
\widetilde{\mathit{C}}_{\alpha,z}(\phi_{S\otimes S})=
\widetilde{\mathit{C}}_{\alpha,z}(\phi_{T\otimes T})=0$.

\vskip0.1in

{\noindent {\bf  Table  1}} Comparisons of the values of
$\mathit{C}_{\alpha,1}(\phi)$ defined in Eq. (\ref{eq8}) with $
\alpha\rightarrow1$ and $\alpha=\frac{1}{2}$ and
$\widetilde{\mathit{C}}_{\alpha,z}(\phi)$ defined in Eq.
(\ref{eq10}). The first column represents the channels,
$p_{\mathrm{min}}$ and $p_{\mathrm{max}}$ represent the values of
$p$ where the maximum and minimum values are attained, respectively.
$\mathrm{max}\lim\limits_{\alpha\rightarrow1}\mathit{C}_{\alpha,1}(\phi)$
and
$\mathrm{min}\lim\limits_{\alpha\rightarrow1}\mathit{C}_{\alpha,1}(\phi)$
represent the maximum and minimum values of
$\lim\limits_{\alpha\rightarrow1}\mathit{C}_{\alpha,1}(\phi)$,
respectively, while $\mathrm{max}\mathit{C}_{\frac{1}{2},1}(\phi)$
and $\mathrm{min}\mathit{C}_{\frac{1}{2},1}(\phi)$ represent the
maximum and minimum values of $\mathit{C}_{\frac{1}{2},1}(\phi)$,
respectively. The last column represents the values of
$\widetilde{\mathit{C}}_{\alpha,z}(\phi)$ defined in Eq.
(\ref{eq10}).
\begin{center}
\renewcommand{\arraystretch}{1.3}
\begin{tabular}{l c c c c c c c c}
\hline
Channels & $p_{\mathrm{max}}$ & $\mathrm{max}\lim\limits_{\alpha\rightarrow1}\mathit{C}_{\alpha,1}(\phi)$ & $p_{\mathrm{min}}$ &$\mathrm{min}\lim\limits_{\alpha\rightarrow1}\mathit{C}_{\alpha,1}(\phi)$ & $\widetilde{\mathit{C}}_{\alpha,z}(\phi)$  \\
\bottomrule

$\phi_{PF}$  &$0$ &$\mathrm{ln2}$ &$\frac{1}{2}$  &$0$ &$0$, $\forall \alpha, z$  \\

$\phi_{D}$  &$0$ &$\mathrm{ln2}$  &$1$  &$0$ &$0$, $\forall \alpha, z$  \\

$\phi_{AD}$  &$0$ &$\mathrm{ln2}$ &$1$ &$0$ &$0$, $\forall \alpha, z$   \\
\hline
Channels & $p_{\mathrm{max}}$ & $\mathrm{max}\mathit{C}_{\frac{1}{2},1}(\phi)$ & $p_{\mathrm{min}}$ &$\mathrm{min}\mathit{C}_{\frac{1}{2},1}(\phi)$ & $\widetilde{\mathit{C}}_{\alpha,z}(\phi)$  \\
\bottomrule

$\phi_{PF}$ &0 &$1$ &$\frac{1}{2}$  &0 &0, $\forall \alpha, z$  \\

$\phi_{D}$ &0 &$1$  &1  &0 &0, $\forall \alpha, z$   \\

$\phi_{AD}$ &0 &1 &1 &0 &0, $\forall \alpha, z$   \\
%$H$  &-  &$ln4$    &-  &$ln4$  &-   \\
%$S$ &-  &$ln2$    &-  &$ln2$  &0, $\forall \alpha, z$    \\
%$T$ &-  &$ln2$    &-  &$ln2$  &0, $\forall \alpha, z$    \\
\hline
\end{tabular}
\end{center}
\vskip0.1in
%$\textbf{Table    2}$ Comparisons the value of $\mathit{C}_{\alpha,1}(\phi)$ defined in Eq. (\ref{eq8}) with $\alpha=\frac{1}{2}$ and $\widetilde{\mathit{C}}_{\alpha,z}(\phi)$ defined in Eq. (\ref{eq10}).
%\begin{center}
%\renewcommand{\arraystretch}{1.5}
%\begin{tabular}{l c c c c c c c}
%\hline
%Channels & $p_{\mathrm{max}}$ & $\mathrm{max}\mathit{C}_{\frac{1}{2},1}(\phi)$ & $p_{\mathrm{min}}$ &$\mathrm{min}\mathit{C}_{\frac{1}{2},1}(\phi)$ & $\widetilde{\mathit{C}}_{\alpha,z}(\phi)$  \\
%\bottomrule
%$\phi_{PF}$ &0 &$1$ &$\frac{1}{2}$  &0 &0, $\forall \alpha, z$  \\
%$\phi_{D}$ &0 &$1$  &1  &0 &0, $\forall \alpha, z$   \\
%$\phi_{AD}$ &0 &1 &1 &0 &0, $\forall \alpha, z$   \\
%$H$ &-  &$\frac{3}{2}$    &-  &$\frac{3}{2}$  &1, $\alpha=\frac{1}{2}, z=1$  \\
%$S$ &-  &1    &-  &1  &0, $\forall \alpha, z$    \\
%$T$ &-  &1    &-  &1  &0, $\forall \alpha, z$   \\
%\hline
%\end{tabular}
%\end{center}
It can be found from Table 1 that under the three quantum channels $\phi_{PF}$, $\phi_{D}$ and $\phi_{AD}$, for either $\alpha\rightarrow1$ or $\alpha=\frac{1}{2}$, $\mathit{C}_{\alpha,1}(\phi)\geq\widetilde{\mathit{C}}_{\alpha,1}(\phi)$ and $\mathit{C}_{\alpha,1}(\phi)$ reaches the maximum value when $p=0$. The minimum values $0$ are attained at the same $p$ for each quantum channel $\phi_{PF}$, $\phi_{D}$ and $\phi_{AD}$. The coherence of $\phi_{PF}$, $\phi_{D}$ and $\phi_{AD}$ have the same maximum values $\mathrm{ln2}$ when $\alpha\rightarrow1$, and the same maximum values $1$ when $\alpha=\frac{1}{2}$.
%For the Hadamard gate $H$, we calculate that $\lim\limits_{\alpha\rightarrow1}\mathit{C}_{\alpha,1}(H)= ln4$
%and $\mathit{C}_{\frac{1}{2},1}(H)=\frac{3}{2}$, which is independent of $p$. For phase gate $S$ and $\frac{\pi}{8}$ gate $T$, we have $\lim\limits_{\alpha\rightarrow1}\mathit{C}_{\alpha,1}(S)=\lim\limits_{\alpha\rightarrow1}\mathit{C}_{\alpha,1}(T)= ln2$, $\mathit{C}_{\frac{1}{2},1}\left(S\right)=\mathit{C}_{\frac{1}{2},1}\left(T\right)=1$ and $\widetilde{\mathit{C}}_{\alpha,z}\left(S\right)=\widetilde{\mathit{C}}_{\alpha,z}\left(T\right)=0$, which are constants independent of $p$.
\vskip0.1in

\noindent {\bf 5. Conclusion}\\\hspace*{\fill}\\
Utilizing the coherence measure of quantum states induced by the generalized $\alpha$-$z$-relative R\'{e}nyi entropy, we have studied the quantifications of the coherence of quantum channels by using two different approaches. Following the idea in \cite{X}, we have introduced a coherence measure of quantum channels by utilizing the Choi-Jamio{\l}kowski isomorphism. We have also verified that $\mathit{C}_{\alpha,z}\left(\phi\right)$ defined in Eq. (\ref{eq7}) is a well-defined coherence measure. On the other hand, inspired by the idea in \cite{FGY}, we have presented an alternative coherence measure by quantifying the commutativity between the channels and the completely dephasing channels with the generalized $\alpha$-$z$-relative R\'{e}nyi entropy. The extremal property, monotonicity and convexity of $\widetilde{\mathit{C}}_{\alpha,z}\left(\phi\right)$ defined in Eq. (\ref{eq10}) have been explored in detail.

Furthermore, the coherence measures defined in Eqs. (\ref{eq8}) and
(\ref{eq10}) have been calculated for some typical channels,
respectively. Analytical formulas of $\mathit{C}_{\alpha,1}(\phi)$
defined in Eq. (\ref{eq8}) for the phase flip channel, depolarizing
channel and amplitude damping channel have been derived and analyzed
for the case of $\alpha\rightarrow1$ and $\alpha=\frac{1}{2}$.
According to Eq. (\ref{eq10}), it can be found that $\phi_{PF}$,
$\phi_{D}$ and $\phi_{AD}$ are all incoherent channels. A table has
been presented to compare different values of coherence measures for
$\phi_{PF}$, $\phi_{D}$ and $\phi_{AD}$. In addition, we have also
considered the unitary channels induced by three quantum gates. The
coherence measures defined in Eqs. (\ref{eq8}) and (\ref{eq10}) for
isotropic channels $\phi_{\Lambda}^H$ with $t\in[-\frac{1}{3},1]$
induced by Hadamard gate have been derived. The quantifiers defined
in Eqs.(\ref{eq8}) and (\ref{eq10}) for unitary channel $\phi_{H}$
induced by Hadamard gate have been deduced as a special case when
$t=1$. The unitary channels induced by $S$ gate and $T$ gate are all
incoherent channels according to Eq. (\ref{eq10}), and they have the
same expressions of $\mathit{C}_{\alpha,1}(\phi)$ as Eq.
(\ref{eq8}). Finally, we have calculated the coherence of quantum
channels induced by $S\otimes S$ and $T\otimes T$ for entangled
qubits, and presented the analytical formulae of the coherence
measures.

Detailed examples and numerical results show that
$\mathit{C}_{\alpha,1}(\phi)\geq
\widetilde{\mathit{C}}_{\alpha,1}(\phi)$ for specific quantum
channels $\phi$, so we conjecture that
$\mathit{C}_{\alpha,1}(\phi)\geq
\widetilde{\mathit{C}}_{\alpha,1}(\phi)$ holds for any quantum
channel, while a rigorous proof is missing. Our results may shed
some new light on the exploration of quantification of coherence for
quantum channels. The regime of coherence quantifiers on the level
of quantum channels needs further study in the future.

\vskip0.1in

\noindent

%=============================================================================%
\subsubsection*{Acknowledgements}
\small {The authors would like to thank the anonymous referees for
their valuable suggestions which greatly improved this paper. This
work was supported by National Natural Science Foundation of China
(Grant Nos. 12161056, 12075159, 12171044); Natural Science
Foundation of Jiangxi Province of China (Grant No. 20232ACB211003);
the Academician Innovation Platform of Hainan Province.}
%===========================================================================%

%=============================================================================%
\subsubsection*{Competing interests}
\small {The authors declare no competing interests.}

%===========================================================================%

%=============================================================================%
\subsubsection*{Data availability}
\small {Data sharing not applicable to this article as no datasets
were generated or analysed during the current study.}

%===========================================================================%

\appendix
\subsubsection* {\bf Appendix A. Calculation of $\mathit{C}_{\alpha,1}(\phi_{PF})$}
According to the Kraus operators of $\phi_{PF}$ given in Example
$1$, we have
\begin{align*}
\mathit{M}_{\phi_{PF}}
&=(\mathbb{I}_2\otimes K_{1})\left(\frac{1}{2}\sum_{i,j=0}^{1}|ii\rangle\langle jj|\right)(\mathbb{I}_2\otimes K_{1})^{\dagger}+(\mathbb{I}_2\otimes K_{2})\left(\frac{1}{2}\sum_{i,j=0}^{1}|ii\rangle\langle jj|\right)(\mathbb{I}_2\otimes K_{2})^{\dagger}\\
&=\frac{1}{2}\left(\begin{array}{cccc} 1 & 0 & 0 & 2p-1\\0 & 0 & 0 &
0\\0 & 0 & 0 & 0\\2p-1 & 0 & 0 & 1\\\end{array}\right),
\end{align*}
where $\mathbb{I}_2$ denotes the $2\times 2$ identity matrix.
Furthermore, we have
\begin{align*}%\label{eq18}
   \mathit{M}_{\phi_{PF}}^{\alpha}
   =&\left(\begin{array}{cccc}
   \frac{p^{\alpha}+(1-p)^{\alpha}}{2} & 0 & 0 & \frac{p^{\alpha}-(1-p)^{\alpha}}{2}\\
   0 & 0 & 0 & 0\\
   0 & 0 & 0 & 0\\
   \frac{p^{\alpha}-(1-p)^{\alpha}}{2} & 0 & 0 & \frac{p^{\alpha}+(1-p)^{\alpha}}{2}
   \end{array}
   \right).
 \end{align*}
Based on $\mathit{M}_{\phi_{PF}}^{\alpha}$, we get
$\mathit{C}_{\alpha,1}(\phi_{PF})$ in Eq. (\ref{eq13}) from Eq.
(\ref{eq8}).

\appendix
\subsubsection* {\bf Appendix B. Calculation of $\mathit{C}_{\alpha,1}(\phi_{D})$}
Direct calculation shows that
\begin{align*}
\mathit{M}_{\phi_{D}}
=&(\mathbb{I}_2\otimes K_{1})\left(\frac{1}{2}\sum_{i,j=0}^{1}|ii\rangle\langle jj|\right)(\mathbb{I}_2\otimes K_{1})^{\dagger}+(\mathbb{I}_2\otimes K_{2})\left(\frac{1}{2}\sum_{i,j=0}^{1}|ii\rangle\langle jj|\right)(\mathbb{I}_2\otimes K_{2})^{\dagger}\\
+&(\mathbb{I}_2\otimes K_{3})\left(\frac{1}{2}\sum_{i,j=0}^{1}|ii\rangle\langle jj|\right)(\mathbb{I}_2\otimes K_{3})^{\dagger}+(\mathbb{I}_2\otimes K_{4})\left(\frac{1}{2}\sum_{i,j=0}^{1}|ii\rangle\langle jj|\right)(\mathbb{I}_2\otimes K_{4})^{\dagger}\\
=&\left(\begin{array}{cccc}
\frac{1}{2}-\frac{p}{4} & 0 & 0 & \frac{1}{2}-\frac{p}{2}\\
0 & \frac{p}{4} & 0 & 0\\
0 & 0 & \frac{p}{4} & 0\\
\frac{1}{2}-\frac{p}{2} & 0 & 0 & \frac{1}{2}-\frac{p}{4}\\
\end{array}\right),
\end{align*}
where $\mathbb{I}_2$ denotes the $2\times 2$ identity matrix. Then
\begin{align*}%\label{eq18}
\mathit{M}_{\phi_{D}}^{\alpha}
=&\left(\begin{array}{cccc}
\frac{1}{2^{2\alpha+1}}p^{\alpha}+\frac{1}{2}
\left(1-\frac{3}{4}p\right)^{\alpha} & 0 & 0 & \frac{1}{2}\left(1-\frac{3}{4}p\right)^{\alpha}-\frac{1}
{2^{2\alpha+1}}p^{\alpha}\\
0 & 4^{-\alpha}p^{\alpha} & 0 & 0\\
0 & 0 & 4^{-\alpha}p^{\alpha} & 0\\
\frac{1}{2}\left(1-\frac{3}{4}p\right)^{\alpha}-\frac{1}
{2^{2\alpha+1}}p^{\alpha} & 0 & 0 & \frac{1}{2^{2\alpha+1}}p^{\alpha}+\frac{1}{2}\left(1-\frac{3}{4}p\right )^{\alpha}
\end{array}
\right),
\end{align*}
from which we get $\mathit{C}_{\alpha,1}(\phi_{D})$ in Eq.
(\ref{eq14}) by using Eq. (\ref{eq8}).

\appendix
\subsubsection* {\bf Appendix C. Calculation of $\mathit{C}_{\alpha,1}(\phi_{AD})$}
According to the Kraus operators of $\phi_{AD}$ given in Example
$3$, we have
\begin{align*}
\mathit{M}_{\phi_{AD}}
=&(\mathbb{I}_2\otimes K_{1})\left(\frac{1}{2}\sum_{i,j=0}^{1}|ii\rangle\langle jj|\right)(\mathbb{I}_2\otimes K_{1})^{\dagger}+(\mathbb{I}_2\otimes K_{2})\left(\frac{1}{2}\sum_{i,j=0}^{1}|ii\rangle\langle jj|\right)(\mathbb{I}_2\otimes K_{2})^{\dagger}\\
=&\frac{1}{2}\left(\begin{array}{cccc} 1 & 0 & 0 & \sqrt{1-p}\\0 & 0
& 0 & 0\\0 & 0 & p & 0\\ \sqrt{1-p} & 0 & 0 &
1-p\\\end{array}\right).
\end{align*}
Then
\begin{align*}
\mathit{M}_{\phi_{AD}}^{\alpha}
=&\left(\begin{array}{cccc}
2^{-\alpha}\left(2-p\right)^{\alpha-1} & 0 & 0 & 2^{-\alpha}\sqrt{\left(1-p\right)}\left(2-p\right)^{\alpha-1}\\
0 & 0 & 0 & 0\\
0 & 0 & 2^{-\alpha}p^{\alpha} & 0\\
2^{-\alpha}\sqrt{\left(1-p\right)}\left(2-p\right)^{\alpha-1} & 0 & 0 & 2^{-\alpha}\left(1-p\right)\left(2-p\right)^{\alpha-1}
\end{array}
\right).
\end{align*}
Utilizing $\mathit{M}_{\phi_{AD}}^{\alpha}$, we derive the formulas
of $\mathit{C}_{\alpha,1}(\phi_{AD})$ in Eq. (\ref{eq15}) via Eq.
(\ref{eq8}).

\appendix
\subsubsection* {\bf Appendix D. Calculation of $\mathit{C}_{\alpha,1}
(\phi_{\Lambda}^{H})$} Noting that
\begin{align*}
\mathit{M}_{\phi_{\Lambda}^{H}}
=&(\mathbb{I}_2\otimes K_{1})\left(\frac{1}{2}\sum_{i,j=0}^{1}|ii\rangle\langle jj|\right)(\mathbb{I}_2\otimes K_{1})^{\dagger}+(\mathbb{I}_2\otimes K_{2})\left(\frac{1}{2}\sum_{i,j=0}^{1}|ii\rangle\langle jj|\right)(\mathbb{I}_2\otimes K_{2})^{\dagger}\\
+&(\mathbb{I}_2\otimes
K_{3})\left(\frac{1}{2}\sum_{i,j=0}^{1}|ii\rangle\langle
jj|\right)(\mathbb{I}_2\otimes K_{3})^{\dagger}+(\mathbb{I}_2\otimes
K_{4})\left(\frac{1}{2}\sum_{i,j=0}^{1}|ii\rangle\langle
jj|\right)(\mathbb{I}_2\otimes
K_{4})^{\dagger}\\+&(\mathbb{I}_2\otimes
K_{5})\left(\frac{1}{2}\sum_{i,j=0}^{1}|ii\rangle\langle
jj|\right)(\mathbb{I}_2\otimes K_{5})^{\dagger}
=\frac{1}{4}\left(\begin{array}{cccc} 1 & t & t & -t\\ t & 1 & t &
-t\\ t & t & 1 & -t\\-t & -t & -t & 1\\\end{array}\right),
\end{align*}
where $\mathbb{I}_2$ denotes the $2\times 2$ identity matrix, we
have \small{
\begin{align*}
\mathit{M}_{\phi_{\Lambda}^{H}}^{\alpha}
=&4^{-1-\alpha}\left(\begin{array}{cccc}
3(1-t)^{\alpha}+(1+3t)^{\alpha} & -(1-t)^{\alpha}+(1+3t)^{\alpha} & -(1-t)^{\alpha}+(1+3t)^{\alpha} & (1-t)^{\alpha}-(1+3t)^{\alpha}\\
-(1-t)^{\alpha}+(1+3t)^{\alpha} & 3(1-t)^{\alpha}+(1+3t)^{\alpha} & -(1-t)^{\alpha}+(1+3t)^{\alpha} & (1-t)^{\alpha}-(1+3t)^{\alpha}\\
-(1-t)^{\alpha}+(1+3t)^{\alpha} & -(1-t)^{\alpha}+(1+3t)^{\alpha} &3(1-t)^{\alpha}+(1+3t)^{\alpha} & (1-t)^{\alpha}-(1+3t)^{\alpha}\\
 (1-t)^{\alpha}-(1+3t)^{\alpha} &(1-t)^{\alpha}-(1+3t)^{\alpha} & (1-t)^{\alpha}-(1+3t)^{\alpha} & 3(1-t)^{\alpha}+(1+3t)^{\alpha}\\\end{array}\right).
\end{align*}}
Making use of $\mathit{M}_{\phi_{\Lambda}^{H}}^{\alpha}$, the
quantity $\mathit{C}_{\alpha,1}(\phi_{\Lambda}^{H})$ in Eq.
(\ref{eq18}) follows immediately from Eq. (\ref{eq8}).

\appendix
\subsubsection* {\bf Appendix E. Calculations of
$\mathit{C}_{\alpha,1}(\phi_{S\otimes S})$ and
$\mathit{C}_{\alpha,1}(\phi_{T\otimes T})$}  Direct calculation
shows that
\begin{align*}
\mathit{M}_{\phi_{S\otimes S}}
=&(\mathbb{I}_4\otimes (S\otimes S))\left(\frac{1}{2}\sum_{i,j=0}^{1}|ii\rangle\langle jj|\otimes \frac{1}{2}\sum_{i,j=0}^{1}|ii\rangle\langle jj|\right)(\mathbb{I}_4\otimes (S\otimes S))^{\dagger}\\
=&\left(\begin{array}{cccccccccccccccc}
\frac{1}{4} & 0 & 0 &-\frac{1}{4} & 0 & 0 & 0 & 0 & 0 & 0 & 0 & 0 & \frac{1}{4} & 0 & 0 & -\frac{1}{4}\\
0 & 0 & 0 & 0 & 0 & 0 & 0 & 0 & 0 & 0 & 0 & 0 & 0 & 0 & 0 & 0\\
0 & 0 & 0 & 0 & 0 & 0 & 0 & 0 & 0 & 0 & 0 & 0 & 0 & 0 & 0 & 0\\
-\frac{1}{4} & 0 & 0 &\frac{1}{4} & 0 & 0 & 0 & 0 & 0 & 0 & 0 & 0 &-\frac{1}{4} & 0 & 0 & \frac{1}{4}\\
0 & 0 & 0 & 0 & 0 & 0 & 0 & 0 & 0 & 0 & 0 & 0 & 0 & 0 & 0 & 0\\
0 & 0 & 0 & 0 & 0 & 0 & 0 & 0 & 0 & 0 & 0 & 0 & 0 & 0 & 0 & 0\\
0 & 0 & 0 & 0 & 0 & 0 & 0 & 0 & 0 & 0 & 0 & 0 & 0 & 0 & 0 & 0\\
0 & 0 & 0 & 0 & 0 & 0 & 0 & 0 & 0 & 0 & 0 & 0 & 0 & 0 & 0 & 0\\
0 & 0 & 0 & 0 & 0 & 0 & 0 & 0 & 0 & 0 & 0 & 0 & 0 & 0 & 0 & 0\\
0 & 0 & 0 & 0 & 0 & 0 & 0 & 0 & 0 & 0 & 0 & 0 & 0 & 0 & 0 & 0\\
0 & 0 & 0 & 0 & 0 & 0 & 0 & 0 & 0 & 0 & 0 & 0 & 0 & 0 & 0 & 0\\
0 & 0 & 0 & 0 & 0 & 0 & 0 & 0 & 0 & 0 & 0 & 0 & 0 & 0 & 0 & 0\\
\frac{1}{4} & 0 & 0 &-\frac{1}{4} & 0 & 0 & 0 & 0 & 0 & 0 & 0 & 0 & \frac{1}{4} & 0 & 0 & -\frac{1}{4}\\
0 & 0 & 0 & 0 & 0 & 0 & 0 & 0 & 0 & 0 & 0 & 0 & 0 & 0 & 0 & 0\\
0 & 0 & 0 & 0 & 0 & 0 & 0 & 0 & 0 & 0 & 0 & 0 & 0 & 0 & 0 & 0\\
-\frac{1}{4} & 0 & 0 &\frac{1}{4} & 0 & 0 & 0 & 0 & 0 & 0 & 0 & 0 &-\frac{1}{4} & 0 & 0 & \frac{1}{4}\\
\end{array}\right)
\end{align*}
and
\begin{align*}
\mathit{M}_{\phi_{T\otimes T}}
=&(\mathbb{I}_4\otimes (T\otimes T))\left(\frac{1}{2}\sum_{i,j=0}^{1}|ii\rangle\langle jj|\otimes \frac{1}{2}\sum_{i,j=0}^{1}|ii\rangle\langle jj|\right)(\mathbb{I}_4\otimes (T\otimes T))^{\dagger}\\
=&\left(\begin{array}{cccccccccccccccc}
\frac{1}{4} & 0 & 0 &\frac{e^{-2\mathrm{i}\pi}}{4} & 0 & 0 & 0 & 0 & 0 & 0 & 0 & 0 & \frac{1}{4} & 0 & 0 & \frac{e^{-2\mathrm{i}\pi}}{4} \\
0 & 0 & 0 & 0 & 0 & 0 & 0 & 0 & 0 & 0 & 0 & 0 & 0 & 0 & 0 & 0\\
0 & 0 & 0 & 0 & 0 & 0 & 0 & 0 & 0 & 0 & 0 & 0 & 0 & 0 & 0 & 0\\
\frac{e^{2\mathrm{i}\pi}}{4}  & 0 & 0 &\frac{1}{4} & 0 & 0 & 0 & 0 & 0 & 0 & 0 & 0 & \frac{e^{2\mathrm{i}\pi}}{4}  & 0 & 0 & \frac{1}{4}\\
0 & 0 & 0 & 0 & 0 & 0 & 0 & 0 & 0 & 0 & 0 & 0 & 0 & 0 & 0 & 0\\
0 & 0 & 0 & 0 & 0 & 0 & 0 & 0 & 0 & 0 & 0 & 0 & 0 & 0 & 0 & 0\\
0 & 0 & 0 & 0 & 0 & 0 & 0 & 0 & 0 & 0 & 0 & 0 & 0 & 0 & 0 & 0\\
0 & 0 & 0 & 0 & 0 & 0 & 0 & 0 & 0 & 0 & 0 & 0 & 0 & 0 & 0 & 0\\
0 & 0 & 0 & 0 & 0 & 0 & 0 & 0 & 0 & 0 & 0 & 0 & 0 & 0 & 0 & 0\\
0 & 0 & 0 & 0 & 0 & 0 & 0 & 0 & 0 & 0 & 0 & 0 & 0 & 0 & 0 & 0\\
0 & 0 & 0 & 0 & 0 & 0 & 0 & 0 & 0 & 0 & 0 & 0 & 0 & 0 & 0 & 0\\
0 & 0 & 0 & 0 & 0 & 0 & 0 & 0 & 0 & 0 & 0 & 0 & 0 & 0 & 0 & 0\\
\frac{1}{4} & 0 & 0 &\frac{e^{-2\mathrm{i}\pi}}{4} & 0 & 0 & 0 & 0 & 0 & 0 & 0 & 0 & \frac{1}{4} & 0 & 0 & \frac{e^{-2\mathrm{i}\pi}}{4} \\
0 & 0 & 0 & 0 & 0 & 0 & 0 & 0 & 0 & 0 & 0 & 0 & 0 & 0 & 0 & 0\\
0 & 0 & 0 & 0 & 0 & 0 & 0 & 0 & 0 & 0 & 0 & 0 & 0 & 0 & 0 & 0\\
\frac{e^{2\mathrm{i}\pi}}{4}  & 0 & 0 &\frac{1}{4} & 0 & 0 & 0 & 0 & 0 & 0 & 0 & 0 & \frac{e^{2\mathrm{i}\pi}}{4}  & 0 & 0 & \frac{1}{4}\\
\end{array}\right),
\end{align*}
where $\mathbb{I}_4$ denotes the $4\times 4$ identity matrix. Then
\begin{align*}
\mathit{M}_{\phi_{S\otimes S}}^{\alpha}
=&\left(\begin{array}{cccccccccccccccc}
\frac{1}{4} & 0 & 0 &-\frac{1}{4} & 0 & 0 & 0 & 0 & 0 & 0 & 0 & 0 & \frac{1}{4} & 0 & 0 & -\frac{1}{4}\\
0 & 0 & 0 & 0 & 0 & 0 & 0 & 0 & 0 & 0 & 0 & 0 & 0 & 0 & 0 & 0\\
0 & 0 & 0 & 0 & 0 & 0 & 0 & 0 & 0 & 0 & 0 & 0 & 0 & 0 & 0 & 0\\
-\frac{1}{4} & 0 & 0 &\frac{1}{4} & 0 & 0 & 0 & 0 & 0 & 0 & 0 & 0 &-\frac{1}{4} & 0 & 0 & \frac{1}{4}\\
0 & 0 & 0 & 0 & 0 & 0 & 0 & 0 & 0 & 0 & 0 & 0 & 0 & 0 & 0 & 0\\
0 & 0 & 0 & 0 & 0 & 0 & 0 & 0 & 0 & 0 & 0 & 0 & 0 & 0 & 0 & 0\\
0 & 0 & 0 & 0 & 0 & 0 & 0 & 0 & 0 & 0 & 0 & 0 & 0 & 0 & 0 & 0\\
0 & 0 & 0 & 0 & 0 & 0 & 0 & 0 & 0 & 0 & 0 & 0 & 0 & 0 & 0 & 0\\
0 & 0 & 0 & 0 & 0 & 0 & 0 & 0 & 0 & 0 & 0 & 0 & 0 & 0 & 0 & 0\\
0 & 0 & 0 & 0 & 0 & 0 & 0 & 0 & 0 & 0 & 0 & 0 & 0 & 0 & 0 & 0\\
0 & 0 & 0 & 0 & 0 & 0 & 0 & 0 & 0 & 0 & 0 & 0 & 0 & 0 & 0 & 0\\
0 & 0 & 0 & 0 & 0 & 0 & 0 & 0 & 0 & 0 & 0 & 0 & 0 & 0 & 0 & 0\\
\frac{1}{4} & 0 & 0 &-\frac{1}{4} & 0 & 0 & 0 & 0 & 0 & 0 & 0 & 0 & \frac{1}{4} & 0 & 0 & -\frac{1}{4}\\
0 & 0 & 0 & 0 & 0 & 0 & 0 & 0 & 0 & 0 & 0 & 0 & 0 & 0 & 0 & 0\\
0 & 0 & 0 & 0 & 0 & 0 & 0 & 0 & 0 & 0 & 0 & 0 & 0 & 0 & 0 & 0\\
-\frac{1}{4} & 0 & 0 &\frac{1}{4} & 0 & 0 & 0 & 0 & 0 & 0 & 0 & 0 &-\frac{1}{4} & 0 & 0 & \frac{1}{4}\\
\end{array}\right)
\end{align*}
and
\begin{align*}
\mathit{M}_{\phi_{T\otimes T}}^{\alpha}
=&\left(\begin{array}{cccccccccccccccc}
\frac{1}{4} & 0 & 0 &\frac{e^{-2\mathrm{i}\pi}}{4} & 0 & 0 & 0 & 0 & 0 & 0 & 0 & 0 & \frac{1}{4} & 0 & 0 & \frac{e^{-2\mathrm{i}\pi}}{4} \\
0 & 0 & 0 & 0 & 0 & 0 & 0 & 0 & 0 & 0 & 0 & 0 & 0 & 0 & 0 & 0\\
0 & 0 & 0 & 0 & 0 & 0 & 0 & 0 & 0 & 0 & 0 & 0 & 0 & 0 & 0 & 0\\
\frac{e^{2\mathrm{i}\pi}}{4}  & 0 & 0 &\frac{1}{4} & 0 & 0 & 0 & 0 & 0 & 0 & 0 & 0 & \frac{e^{2\mathrm{i}\pi}}{4}  & 0 & 0 & \frac{1}{4}\\
0 & 0 & 0 & 0 & 0 & 0 & 0 & 0 & 0 & 0 & 0 & 0 & 0 & 0 & 0 & 0\\
0 & 0 & 0 & 0 & 0 & 0 & 0 & 0 & 0 & 0 & 0 & 0 & 0 & 0 & 0 & 0\\
0 & 0 & 0 & 0 & 0 & 0 & 0 & 0 & 0 & 0 & 0 & 0 & 0 & 0 & 0 & 0\\
0 & 0 & 0 & 0 & 0 & 0 & 0 & 0 & 0 & 0 & 0 & 0 & 0 & 0 & 0 & 0\\
0 & 0 & 0 & 0 & 0 & 0 & 0 & 0 & 0 & 0 & 0 & 0 & 0 & 0 & 0 & 0\\
0 & 0 & 0 & 0 & 0 & 0 & 0 & 0 & 0 & 0 & 0 & 0 & 0 & 0 & 0 & 0\\
0 & 0 & 0 & 0 & 0 & 0 & 0 & 0 & 0 & 0 & 0 & 0 & 0 & 0 & 0 & 0\\
0 & 0 & 0 & 0 & 0 & 0 & 0 & 0 & 0 & 0 & 0 & 0 & 0 & 0 & 0 & 0\\
\frac{1}{4} & 0 & 0 &\frac{e^{-2\mathrm{i}\pi}}{4} & 0 & 0 & 0 & 0 & 0 & 0 & 0 & 0 & \frac{1}{4} & 0 & 0 & \frac{e^{-2\mathrm{i}\pi}}{4} \\
0 & 0 & 0 & 0 & 0 & 0 & 0 & 0 & 0 & 0 & 0 & 0 & 0 & 0 & 0 & 0\\
0 & 0 & 0 & 0 & 0 & 0 & 0 & 0 & 0 & 0 & 0 & 0 & 0 & 0 & 0 & 0\\
\frac{e^{2\mathrm{i}\pi}}{4}  & 0 & 0 &\frac{1}{4} & 0 & 0 & 0 & 0 & 0 & 0 & 0 & 0 & \frac{e^{2\mathrm{i}\pi}}{4}  & 0 & 0 & \frac{1}{4}\\
\end{array}\right),
\end{align*}
By Eq. (\ref{eq8}), we can thus deduce
$\mathit{C}_{\alpha,1}(\phi_{S\otimes S})$ and
$\mathit{C}_{\alpha,1}(\phi_{T\otimes T})$ in Eq. (\ref{eq23}) based
on $\mathit{M}_{\phi_{S\otimes S}}^{\alpha}$ and
$\mathit{M}_{\phi_{T\otimes T}}^{\alpha}$.

%===========================================================================%

\end{sloppypar}
\end{document}